\documentclass[JCAP]{article} 
\usepackage{jcappub}
\usepackage{times}

\usepackage{epstopdf}
\usepackage{graphicx}
\usepackage{dcolumn}
\usepackage{bm}
\usepackage{color}
\usepackage{hyperref}
\usepackage{multirow}
\input epsf

\begin{document}
\title{Non-linear evolution of the cosmic neutrino background}

\author[a]{Francisco Villaescusa-Navarro,} \author[b]{Simeon Bird,} \author[c]{Carlos Pe\~na-Garay,} \author[a,d]{Matteo Viel} 

\affiliation[a]{INAF/Osservatorio Astronomico di Trieste, Via Tiepolo 11, 34143, Trieste, Italy}
\affiliation[b]{Institute for Advanced Study, 1 Einstein Drive, Princeton, NJ, 08540, USA}
\affiliation[c]{Instituto de F\'isica Corpuscular, CSIC-UVEG, E-46071, Paterna, Valencia, Spain}
\affiliation[d]{INFN sez. Trieste, Via Valerio 2, 34127 Trieste, Italy}

\emailAdd{villaescusa@oats.inaf.it}
\emailAdd{spb@ias.edu}
\emailAdd{penya@ific.uv.es}
\emailAdd{viel@oats.inaf.it}

\abstract{We investigate the non-linear evolution of the relic cosmic
  neutrino background by running large box-size, high resolution
  N-body simulations which incorporate cold dark matter (CDM) and
  neutrinos as independent particle species. Our set of simulations
  explore the properties of neutrinos in a reference $\Lambda$CDM
  model with total neutrino masses between 0.05-0.60 eV in cold dark
  matter haloes of mass $10^{11}-10^{15}$ $h^{-1}$M$_{\odot}$, over a
  redshift range $z=0-2$. We compute the halo mass function and show
  that it is reasonably well fitted by the Sheth-Tormen formula, once
  the neutrino contribution to the total matter is removed.  More
  importantly, we focus on the CDM and neutrino properties of the
  density and peculiar velocity fields in the cosmological volume,
  inside and in the outskirts of virialized haloes.  The dynamical
  state of the neutrino particles depends strongly on their momentum:
  whereas neutrinos in the low velocity tail behave similarly to CDM
  particles, neutrinos in the high velocity tail are not affected by
  the clustering of the underlying CDM component. We find that the
  neutrino (linear) unperturbed momentum distribution is modified and
  mass and redshift dependent deviations from the expected Fermi-Dirac
  distribution are in place both in the cosmological volume and inside
  haloes.  The neutrino density profiles around virialized haloes have
  been carefully investigated and a simple fitting formula is
  provided. The neutrino profile, unlike the cold dark matter one, is
  found to be cored with core size and central density that depend on
  the neutrino mass, redshift and mass of the halo, for halos of
  masses larger than $\sim 10^{13.5}h^{-1}$M$_{\odot}$. For lower
  masses the neutrino profile is best fitted by a simple power-law
  relation in the range probed by the simulations.  The results we
  obtain are numerically converged in terms of neutrino profiles at
  the 10\% level for scales above $\sim 200$ $h^{-1}$kpc at $z=0$, and
  are stable with respect to box-size and starting redshift of the
  simulation.  Our findings are particularly important in view of
  upcoming large-scale structure surveys, like Euclid, that are
  expected to probe the non-linear regime at the percent level with
  lensing and clustering observations.}

\maketitle

\section{Introduction}

The standard Big Bang cosmology predicts the existence of a relic
particle radiation in the form of neutrinos, the cosmic neutrino
background. In the primordial plasma, the three standard model
neutrino flavors are in thermal equilibrium with photons, electrons
and positrons and their momentum distribution followed the Fermi-Dirac
distribution. When the temperature of the universe drops to
$\sim10^{10}$K$~(\sim 1~\rm{MeV})$, the expansion rate becomes larger than
the rate of neutrino interactions, triggering the decoupling of the
neutrinos from the rest of the plasma
\cite{KolbTurner,Dodelson,LesgourguesPastor}. At the time of neutrino
decoupling, neutrino momenta are much larger than their masses, and
thus, the number density of neutrinos with momentum between $p$ and
$p+dp$ is:
\begin{equation}
n(p)dp=\frac{4\pi g_\nu}{(2\pi\hbar
c)^3}\frac{p^2dp}{e^\frac{p}{k_BT_\nu}+1}~,
\label{eq1}
\end{equation}
where $g_\nu$ is the number of neutrino spin states, $k_B$ is the
Boltzmann constant and $T_\nu=T_\nu(z_{dec})$ is the temperature of
the universe at the neutrino decoupling time {\footnote{Note that we
    have made the approximation
    $E_\nu(z_{dec})=\sqrt{m_\nu^2+p_\nu^2(z_{dec})}\cong
    p_\nu(z_{dec})$.}}. Once neutrinos decouple, their momentum is
redshifted as $1/(1+z)$ and their distribution is given by
Eq. \ref{eq1} with
$T_\nu=T_\nu(z)=T_\nu(z_{dec})(1+z)/(1+z_{dec})$. It can be shown (see
for example \cite{Weinberg}) that current neutrino and CMB
temperatures are related through
$T_\nu(z=0)=\left(\frac{4}{11}\right)^{1/3}T_\gamma(z=0)$ (a small
correction to this formula arises when taking into account that some
neutrinos are still weakly coupled at the electron-positron
annihilation \cite{Mangano}).  Equation \ref{eq1} represents the
unperturbed momentum distribution of the cosmic neutrino background,
i.e. in deriving it we have considered only the redshift of neutrino
momentum due to the expansion of the universe. The number density and
mean thermal velocity of cosmic neutrinos (neutrinos plus
antineutrinos) can be computed from equation \ref{eq1}, resulting in
$\overline{n}_\nu(z)\cong113(1+z)^3\frac{\nu}{\rm{cm}^3}$ and
$\overline{V}_\nu(z)\cong160(1+z)\left(\frac{\rm{eV}}{m_\nu}\right)~\rm{km/s}$,
respectively.

We have indirect evidence of the existence of the relic neutrino
background. The three light neutrino species contribute to the
radiation energy density, changing the expansion rate of the Universe
and the time of matter-radiation equality. This in turn affects both
Big Bang nucleosynthesis, and thus the primordial abundances of light
elements, and the cosmic microwave background (CMB)
anisotropies. Finally, late-time free-streaming of the neutrinos
suppresses the growth of matter perturbations
\cite{KolbTurner,Dodelson,LesgourguesPastor}. This effect is well
understood at the linear level, but on non-linear scales other
theoretical or numerical tools have to be used, such as N-body
simulations
\cite{MaBertschinger,Valdarnini,Brandbyge2008,Brandbyge2009,BrandbygeHybrid,Viel_2010,
  Agarwal2011, Bird_2011,Wagner2012, AliHaimoudBird}, perturbation
theory \cite{Saito,LesgourguesMatarrese} or semi-analytical methods
\cite{Ma,Wong,Paco,Abazajian,Ichiki,Shoji}.

From the particle physics side, the discovery of flavor conversion in
neutrino experiments, coined neutrino oscillations, implies that at
least two of the three neutrino species are massive, with minimal
masses of about 9 and 50 meV \cite{Concha}. From the cosmological
side, an important and competitive way to constrain the mass and
number of cosmological neutrinos is offered by Large-Scale Structure
(LSS) data. The clustering properties of cold dark matter (CDM) and
neutrinos (hot dark matter) are very different. Neutrino clustering is
strongly influenced by the clustering of the dominant CDM component,
but the much larger thermal velocities of neutrinos, as compared to
the CDM one, leads to a suppression of the total clustering on small
scales and could produce neutrino overdensities in regions where the
CDM density is high, in a mass and redshift dependent fashion.

Present cosmological data put an upper limit of $\sim 0.30$ eV, at the
$2\sigma$ confidence level, on the total neutrino mass by using large
LSS data such as SDSS\footnote{http://www.sdss.org/} luminous red
galaxies, CFHTLS \footnote{http://www.cfht.hawaii.edu/Science/CFHLS/}
or WiggleZ \footnote{http://wigglez.swin.edu.au/site/} (e.g.~
\cite{Reid, Thomas, Swanson, dePutter, Xia2012,WiggleZ, Zhao2012}).
With the notable exceptions of the work of \cite{Xia2012}, based on CFHTLS and
VIPERS \footnote{http://vipers.inaf.it/} galaxy clustering data, and
the analysis of \cite{Seljak2006}, that rely on the Lyman-$\alpha$ forest ($\Sigma_i
m_{\nu_i}<0.17$ eV), these constraints use only the information
contained in the linear regime.

Therefore, the sum of neutrino masses is constrained within an order
of magnitude and the model with massless neutrinos has to be modified
by a specific model of massive neutrinos with two hierarchical mass
splittings.  Incoming data analysis of the CMB \cite{Story, Ade} will
further improve the significance on the existence of relic neutrinos
and with the LSS data in the linear regime may sign to a positive
signal on the total mass of neutrinos
(e.g. \cite{Jimenez,Takada,Hamann,Audren}). These cosmological
observations are thereby very important. Though we do not have a
definitive theory of flavor, the total mass determination would allow
to distinguish between two large groups of models: degenerate vs
hierarchical mass models. More importantly, a large total neutrino
mass (in the sensitivity range of running CMB and LSS observations),
will fix the predictions of majorana neutrinos in the minimal
extension of the standard model of particles physics, within the range
of sensitivity of next generation neutrinoless double beta decay
experiments \cite{GomezCadenas}, and therefore, would allow the
experiment to distinguish whether neutrinos are their own
antiparticles, and unambiguously imply the existence of a new high
energy scale in physics \cite{Kayser}.

While some recent works have explored the potential of observables
sensitive to the neutrino mass in the nonlinear regime
\cite{Wong,Paco} by other methods, high resolution large box-size
N-body simulations are the most accurate way to describe non-linear
gravitational clustering. Previous works using N-body simulations
including neutrinos have examined the effect on the matter power
spectrum \cite{Brandbyge2008,Viel_2010,Bird_2011,Wagner2012}.  In this
work, we will address more closely the density and peculiar velocity
fields, similarly to \cite{Brandbyge_haloes} but with improved
simulations and a more extensive analysis that brackets a large
dynamical and redshift range.  We will study the evolution of the
non-linear distribution of neutrinos in the whole simulated box, as
well as the neutrino properties inside and around virialized
haloes. In particular, we will compute and characterize the neutrino
density profile around massive CDM halos, providing the reader a
fitting formula that reproduces them. Also, we will compute the
neutrino momentum distribution and determine how closely it adheres to
equation \ref{eq1}.

The paper is organized as follows. In Section \ref{Simulations} we
describe our N-body simulations.  The evolution of the neutrinos and
dark matter density and peculiar velocity fields, the halo mass
functions and the properties inside and in the neighborhood of dark
matter haloes will be addressed in Section \ref{Results}.  This
section will contain the main results of the paper and present fitting
formula for several different quantities (in the appendix we will show
the dependence of some of the physical quantities on the
environment). Finally, in Section \ref{Conclusions}, we summarize the
main results of this work and present future perspectives.

\section{The simulations}
\label{Simulations}

We use particle-based N-body simulations containing CDM and neutrino
particles, performed using a modified version of the TreePM code {\sc
  GADGET}-3, as described in detail in \cite{Springel_2005,Viel_2010,
  Bird_2011}.  Neutrinos are treated as dark matter particles, with a
large initial thermal velocity drawn from the Fermi-Dirac distribution
of Eq. \ref{eq1}.  The time-step used by the code is set by the CDM
only, and is not affected by the neutrino particles.  The force on the
neutrinos includes the contribution from the short-range tree: this is
in contrast to \cite{Viel_2010}, where the neutrino force was computed
using the long-range particle-mesh only.  We found that the
short-range tree force is required to properly resolve the clustering
of the neutrinos in the center of massive halos at low redshift.
However, for simulation S30 only (see text below), for reasons of
performance the tree is disabled for the neutrinos between $z=99$ and
$z=20$.

\begin{table}
\begin{center}
\begin{tabular}{|c|c|c|c|c|c|c|}

\hline
Name & $\Sigma_i m_{\nu_i}$ (eV) & Box ($h^{-1}\rm{Mpc}$) & $N_\mathrm{DM}^{1/3}$ & $N_\nu^{1/3}$ & $z_i$ & $\sigma_8$ ($z=0$) \\
\hline
\hline 
H60 & $0.60$ & $1000$ & $512$ & $1024$ & $19$ & $0.6760$ \\
\hline
L60 & $0.60$ & $500$ & $512$ & $1024$ & $99$ & $0.6760$ \\
\hline
L45 & $0.45$ & $500$ & $512$ & $1024$ & $99$ & $0.7133$ \\
\hline
L30 & $0.30$ & $500$ & $512$ & $1024$ & $99$ & $0.7531$ \\
\hline
L15 & $0.15$ & $500$ & $512$ & $1024$ & $99$ & $0.7947$ \\
\hline
L0 & $0.00$ & $500$ & $512$ & $0$ & $99$ & $0.8325$ \\
\hline
S60 & $0.60$ & $100$ & $512$ & $1024$ & $99$ & $0.6760$ \\
\hline
S30 & $0.30$ & $100$ & $512$ & $1024$ & $99$ & $0.7531$ \\
\hline
LL60 & $0.60$ & $500$ & $512$ & $512$ & $99$ & $0.6760$ \\
\hline
LL45 & $0.45$ & $500$ & $512$ & $512$ & $99$ & $0.7133$ \\
\hline
LL5 & $0.05$ & $500$ & $512$ & $512$ & $99$ & $0.8120$ \\
\hline

\end{tabular}
\end{center} 
\caption{Summary of simulation parameters. $m_{\nu_i}$ is the mass of a 
single neutrino species, and $\Sigma_i m_{\nu_i}$ the total neutrino mass.
Cosmological parameters are the same for all simulations and are given
in the text. $\Omega_{\rm M}$ is kept constant, so that for an increase in 
neutrino mass $\Omega_\mathrm{cdm}$ decreased and the neutrinos make 
up an increased fraction of the total dark matter.}
\label{tab_sims}
\end{table}

Our initial conditions are produced from transfer functions generated
by \textsc{camb} \cite{CAMB}, using our own version of N-GenICs
modified to use second order Lagrangian perturbation theory
\cite{Scoccimarro_1998} for the CDM particles\footnote{Our initial
  conditions generator is freely available at
  \url{http://github.com/sbird/S-GenIC}.}.  The transfer function used
for our cold dark matter (CDM) particles is a weighted average of the
linear theory transfer functions for CDM and baryons, to account for
the slight difference between them \citep{Somogyi_2010}.  For the
neutrinos initial conditions we use the Zel'dovich approximation
\citep{Zeldovich_1970}, with identical initial random phase
information to the CDM one, in order to simulate adiabatic initial
conditions.  Our cosmological parameters are the following:
$\Omega_{\rm{b}}=0.05$,
$\Omega_{\rm{M}}=\Omega_{\rm{CDM}}+\Omega_{\rm{b}}+\Omega_{\nu}=0.2708$,
$\Omega_\Lambda=0.7292$, $n_{\rm{s}}=1.0$, $h=0.703$,
$A_{\rm{s}}=2.43\times10^{-9}$, which are roughly in agreement with
\cite{WMAP7}. The massless neutrino case has a $\sigma_8=0.8325$ which
is also in reasonable agreement with LSS data.  The starting redshift
of most of our simulations is set $z=99$. However, we check that our
results are insensitive to this value with simulations started at
later times ($z=49$ and $z=19$).  We used a variety of box sizes and
particle numbers to ensure our results were independent of both
unresolved large-scale modes and unresolved small-scale structure.  In
particular we simulate three different boxes of linear size 1000, 500
and 100 com. $h^{-1}$ Mpc with always the same number of CDM
particles ($512^3$). For the neutrinos we use two different numbers of
total neutrino particles in order to address the neutrino shot-noise
($512^3$ and $1024^3$) and simulate five different realizations, each of them corresponding to three degenerate neutrino species with $\Sigma_i m_{\nu_i}=0.15,~0.30,~0.45,~0.60$ eV. We also consider the case with $\Sigma_i m_{\nu_i}=0.05$ eV, for which we produce an initial power spectrum considering only one massive neutrino species.

The gravitational softening length is set to $1/40$ of the mean
inter-particle spacing for the neutrinos, and the number of cells per
side in the particle-mesh grid is $N_\nu^{1/3}$.  The parameters of
our simulations are shown in Table \ref{tab_sims}.  We also run a
simulation identical to L60, but with a different random seed for the
initial structure field, to verify that our results were insensitive
to the realisation of cosmic structure. The total CPU time consumption
for the neutrino simulations are between 15000-45000 hrs.

\section{Results}
\label{Results}

In this Section we analyse the suite of simulations and present our
main results\footnote{Several movies showing the distributions of CDM and $\nu$ are available at \url{http://som.ific.uv.es/movies}}. Firsly, we will focus on non-linear properties on large
scales, studying the probability distribution function of both the
non-linear density field and the non-linear peculiar velocity
field. Then we will consider dark matter haloes and compute the mass
function for different cosmologies with different neutrino masses at
several redshifts.  We will also study in detail the clustering of
cosmological relic neutrinos within the gravitational potential wells
of CDM haloes, providing the reader with a fitting function for the
neutrino density profiles.

\subsection{The non-linear density field}

\begin{figure}
\begin{center}
\includegraphics[width=1.0\textwidth]{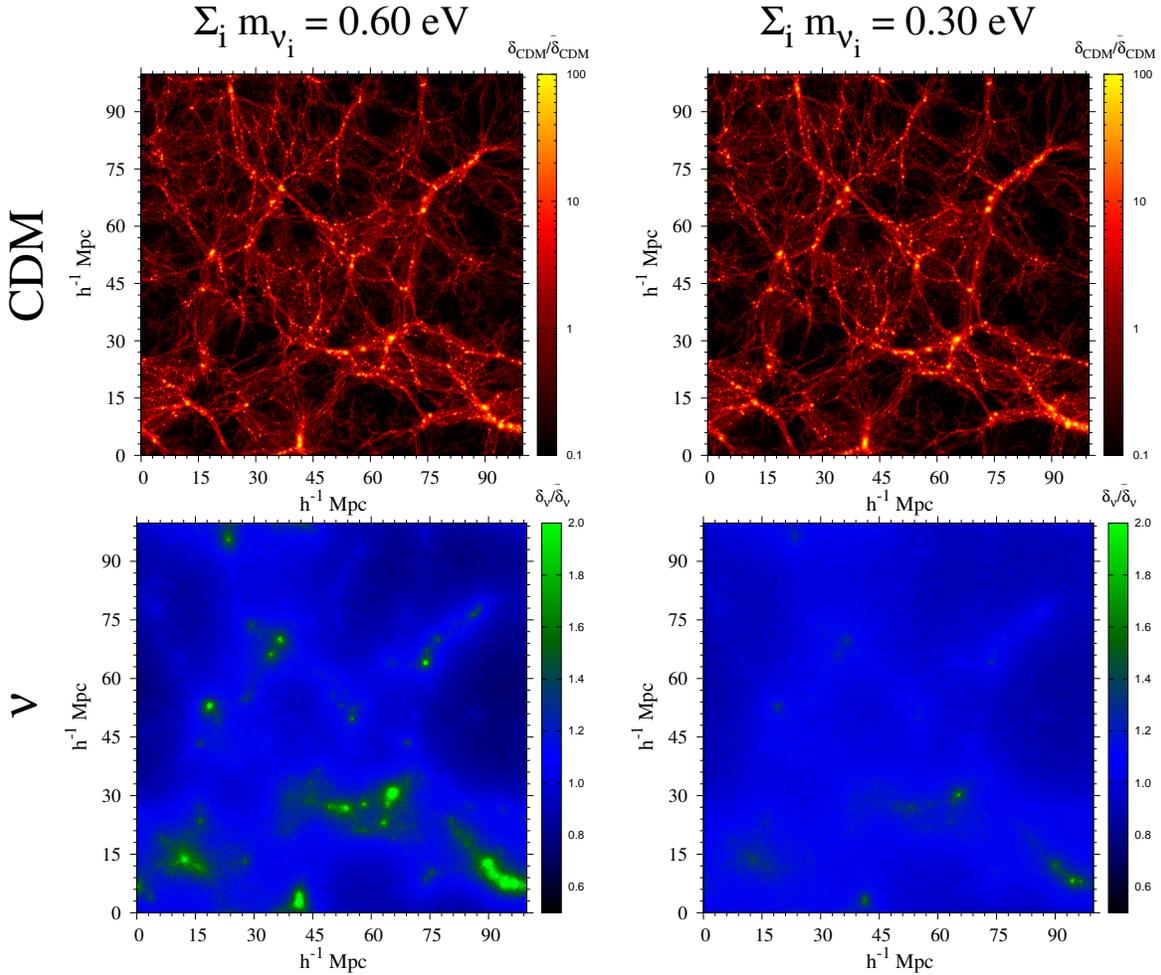}
\end{center}
\caption{Slice of thickness 5 com. $h^{-1}$ Mpc through the density
  field of CDM and neutrinos. The upper panels show a slice of the CDM
  density field extracted from a N-body simulation with neutrinos of
  masses $\Sigma_i m_{\nu_i}=0.60$ eV (left column) and $\Sigma_i
  m_{\nu_i}=0.30$ eV (right column). The bottom panels show the
  neutrino density field in the same slices of the upper panels.}
\label{Slice}
\end{figure}

\begin{figure}
\begin{center}
\includegraphics[width=1.0\textwidth]{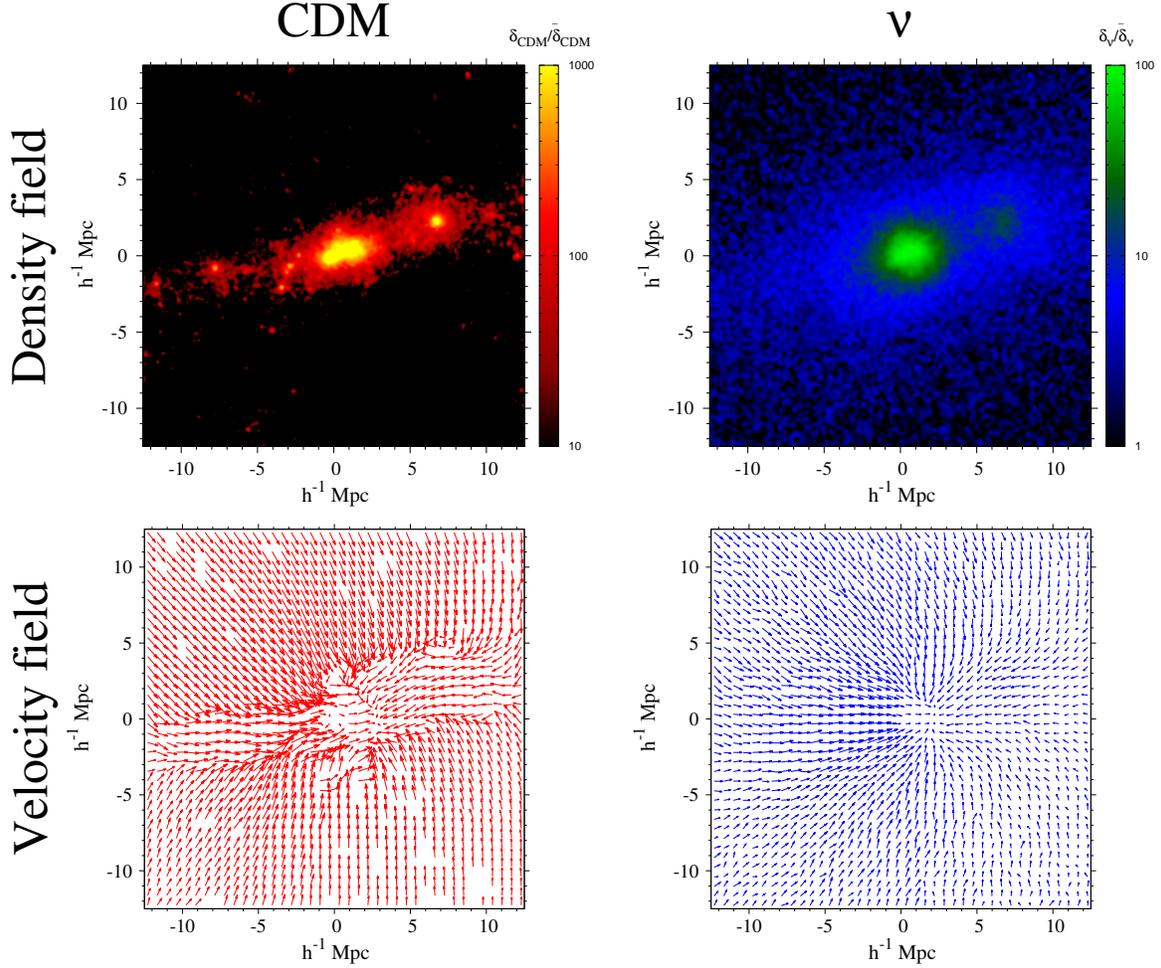}
\end{center}
\caption{Distribution of densities and peculiar velocities around the
  most massive CDM halo as extracted from the simulation L60. The
  position of the center of the halo is (0,0) $h^{-1}$Mpc in all
  panels. \textit{Upper left:} CDM density distribution. \textit{Upper
    right:} neutrino density distribution. \textit{Bottom left:} CDM
  peculiar velocities. \textit{Bottom right:} neutrino peculiar
  velocities. In all the panels the quantities have been projected
  over the same plane. All figures have been computed using the CIC
  interpolation scheme in a slice of linear size $25$ $h^{-1}$Mpc having a
  thickness of 4 $h^{-1}$Mpc (comoving units).}
\label{FoM}
\end{figure}

In Fig. \ref{Slice} we show a slice of thickness 5 com. $h^{-1}$ Mpc
for the density fields of the CDM component and neutrinos of masses
$\Sigma_i m_{\nu_i}=0.60$ eV (left panels) and $\Sigma_i
m_{\nu_i}=0.30$ eV (right panels). As expected, the regions around
which neutrinos tend to cluster are those where the density of CDM is
large. This can be seen even more clearly in Fig. \ref{FoM}, where we
show the density and peculiar velocity fields, for the CDM and the
cosmological neutrinos, in the neighborhood of the most massive halo
present in simulation L60. As we shall see, the neutrino clustering
depends on the total neutrino mass, the mass of the haloes and
redshift.

\begin{figure}
\centering
\includegraphics[width=0.495\textwidth]{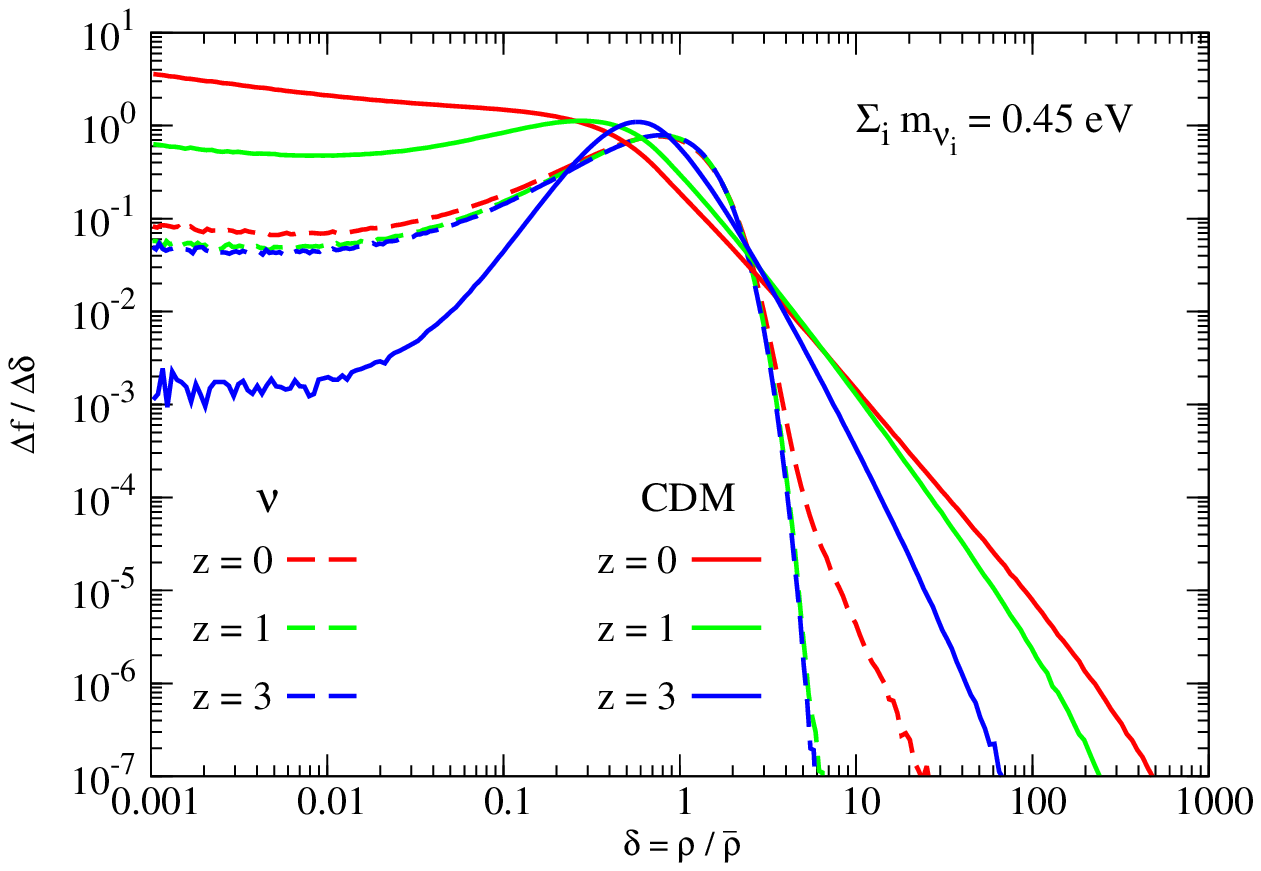}
\includegraphics[width=0.495\textwidth]{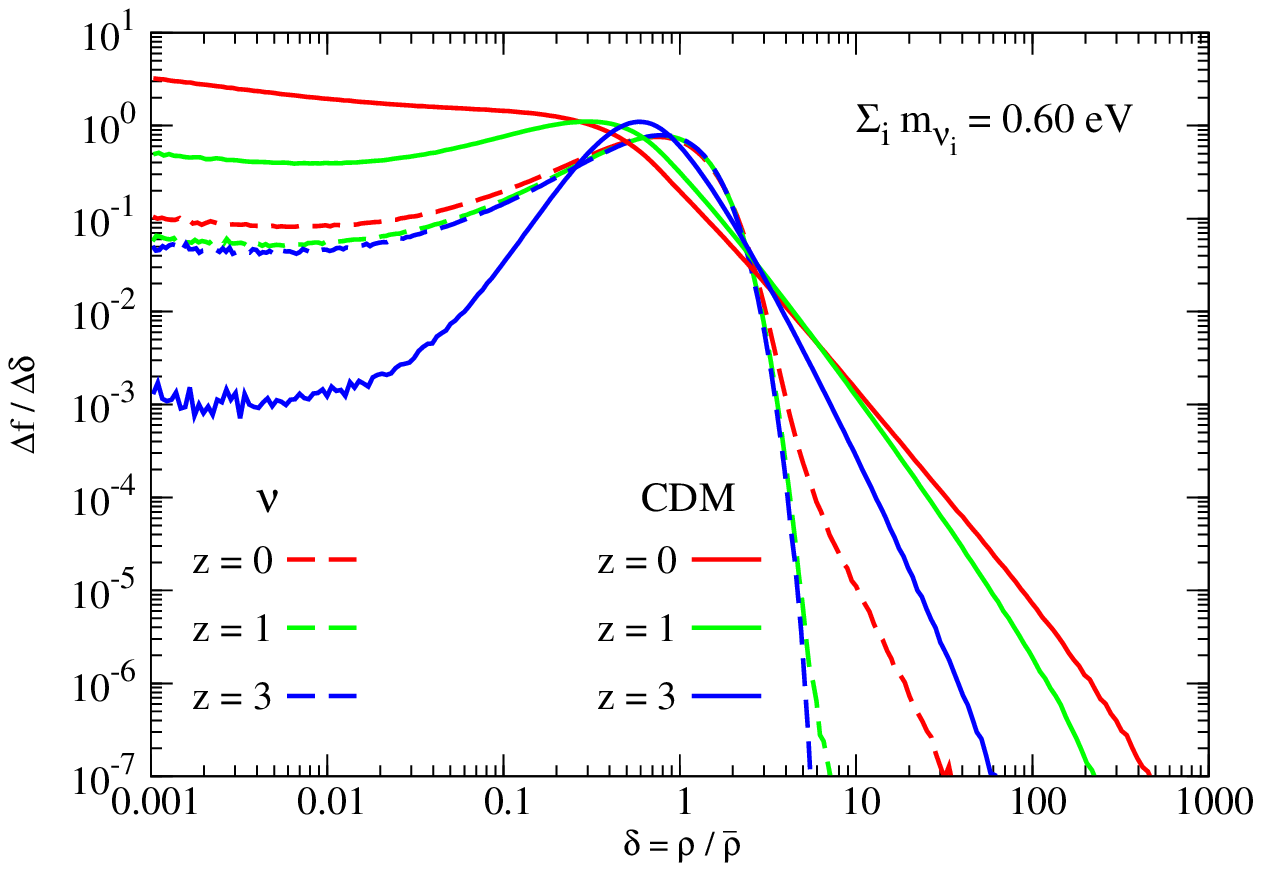}
\caption{Redshift evolution of the non-linear density field computed
  over the whole cosmological simulated volume. We use the CIC
  interpolation scheme to compute the value of density field, for CDM
  and neutrinos, in a regular grid of $500\times500\times500$
  points. We show the fraction of grid points with overdensities
  between $\delta$ and $\delta+\triangle\delta$, per $\triangle
  \delta$, at three different redshifts $z=0,1,3$ (red, green and blue
  curves, respectively) for CDM (solid lines) and neutrinos (dashed
  lines). The left panel shows the results for the simulation LL45,
  while the right panel corresponds to the simulation LL60.}
\label{CIC_density}
\end{figure}

We now turn to a more detailed analysis of the density field by
computing the value of the density field in a grid of
$500\times500\times500$ points using the cloud-in-cell (CIC)
interpolation. This is done for the density fields of both the CDM and the
neutrinos. The simulations we have used are LL45 and LL60 (which
have the same number of CDM and neutrino particles). In Figure
\ref{CIC_density} we plot, as a function of the overdensity, 
$\delta=\rho/\overline{\rho}$, the fraction of grid points whose
overdensities are between $\delta$ and $\delta+\triangle\delta$, per
$\triangle \delta$, at three different redshifts: $z=0,1,3$. The
results for the cosmological models with $\Sigma_i m_{\nu_i}=0.45$ eV
and $\Sigma_i m_{\nu_i}=0.60$ eV are plotted on the left and right
panels, respectively.

As envisaged, the distribution of the CDM density field stretches out
at lower redshift as the cosmic structure form: voids become emptier
at the expenses of overdense regions that become denser and denser
with decreasing redshift. We note that the distributions for the CDM
component are not exactly equal in the two cosmological models. In the
cosmological model with neutrino masses equal to $\Sigma_i
m_{\nu_i}=0.45$ eV (left panel), the number of grid points with high
or low overdensities is slightly larger than in the cosmology with
$\Sigma_i m_{\nu_i}=0.60$ eV. This is due to their different power
spectra, as reflected in the value of $\sigma_8=0.7133$ (0.45 eV) and
$\sigma_8=0.676$ (0.60 eV). In contrast, the distribution of neutrinos
displays a much slower evolution than CDM: while the fraction of grid
points with low CDM densities grows rapidly, the fraction of grid
points with low neutrino densities barely changes with time. On the
other hand, the fraction of grid points with large neutrino
overdensities increases significantly between $z=1$ and $z=0$,
reflecting the fact that non-linear neutrino clustering takes place
only at low redshift, as also found by \cite{AliHaimoudBird}. We also
find that the neutrino density field evolves more slowly for lower
neutrino masses: this happens because lighter neutrinos have higher
thermal velocities which prevents them from clustering into CDM haloes
or being excluded from the interior of cosmological voids. It is
important to remark that the results we show in Fig. \ref{CIC_density}
are not fully numerically converged. By repeating the same procedure
for the simulations L45 and L60 we find that only the high density
tail, $\delta\gtrsim 5$, is converged. The aim of a calculation like
the one presented in this Section is to qualitatively show the reader
the different redshift evolution of the CDM and cosmic neutrino
density fields.

\subsection{The non-linear velocity field}
\label{nl_velocity_field}

We now focus on the time evolution of the non-linear peculiar velocity
field of both CDM and neutrinos. As in the case of the non-linear
density field, we expect the velocity fields of CDM and neutrinos to
behave differently. On one hand, by definition, the CDM has negligible
peculiar velocities initially. We would thus expect that on average,
the CDM particles will increase their peculiar velocities since, among
others, they will cluster within virialized haloes and escape from
cosmological voids. On the other hand, neutrinos should behave in the
opposite way: at high redshift neutrinos have very large thermal
velocities, and if we neglect effects associated with gravity, such as
neutrino clustering, their thermal velocities should drop with time as
$1/(1+z)$.

\begin{figure}
\centering
\includegraphics[width=0.495\textwidth]{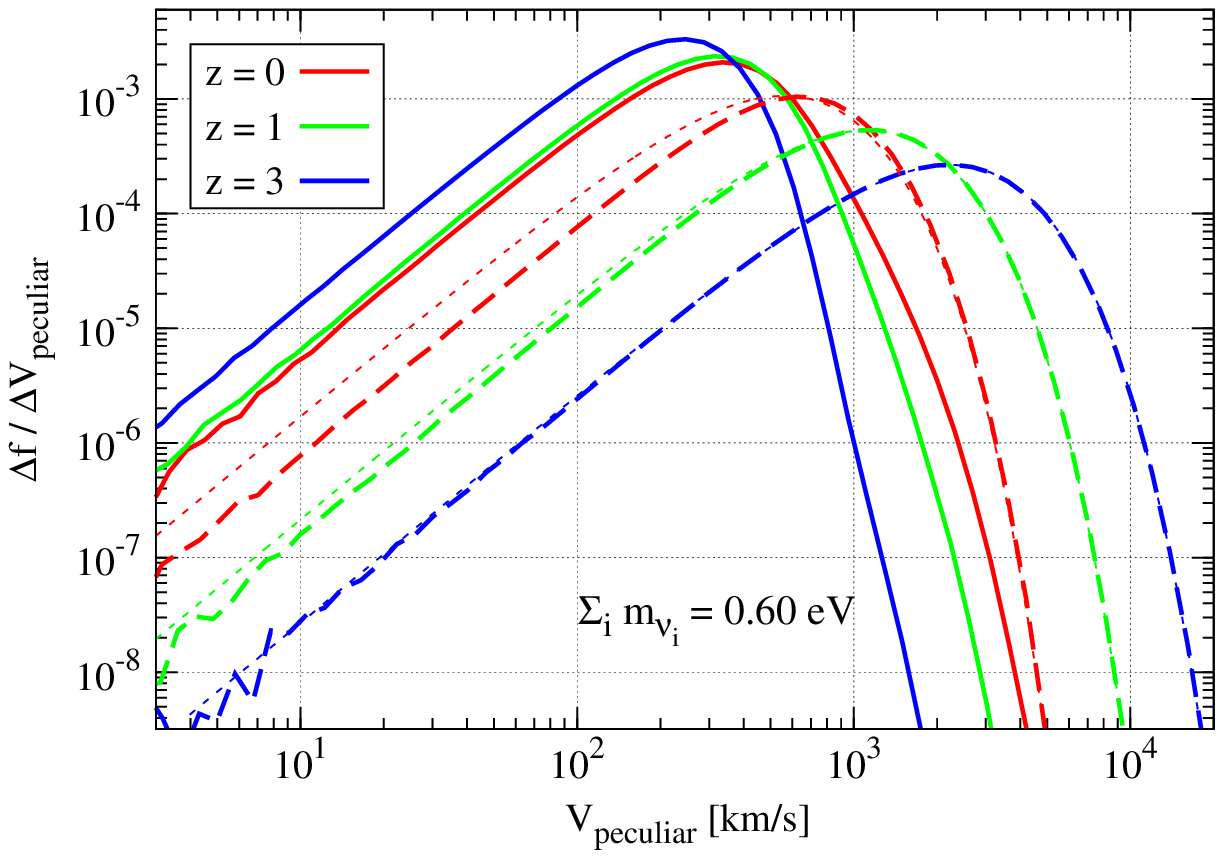}
\includegraphics[width=0.495\textwidth]{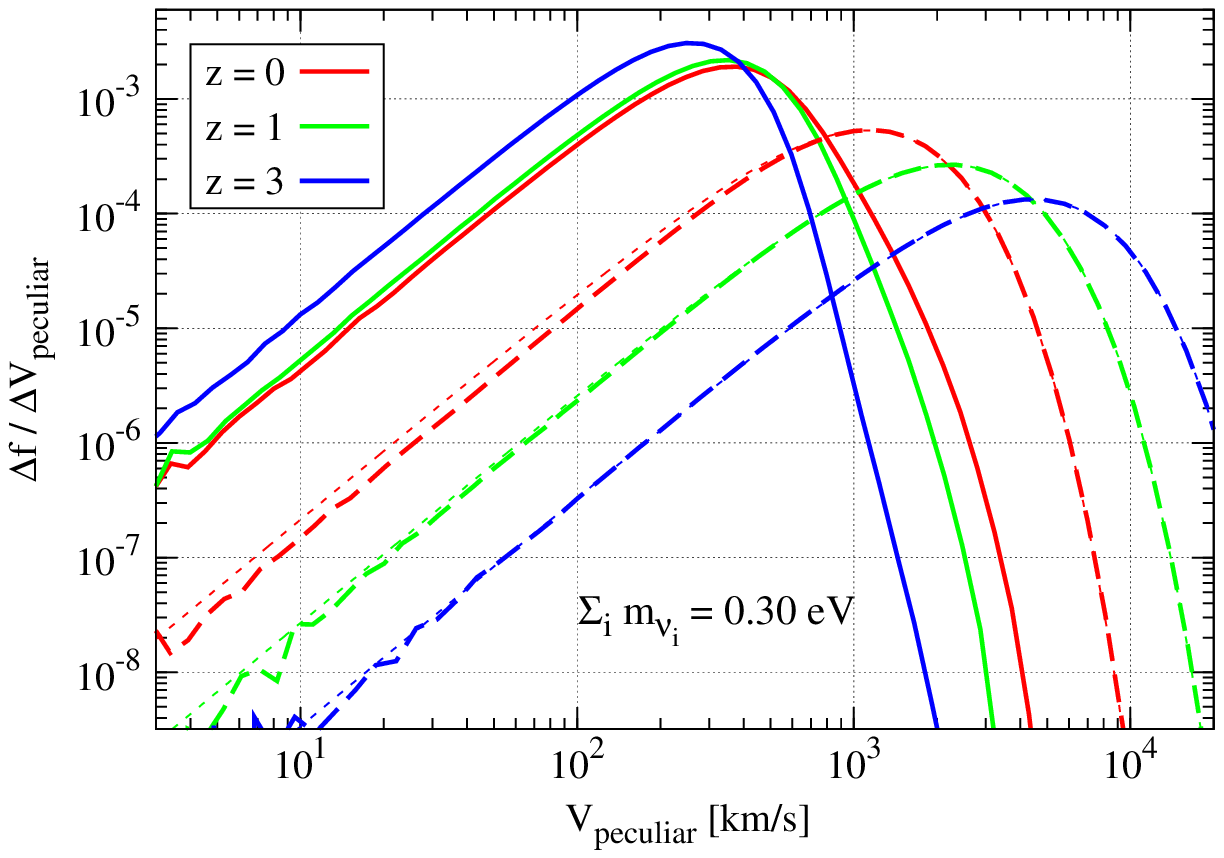}
\caption{Fraction of particles with peculiar velocities between $V$
  and $V+\bigtriangleup V$, per $\bigtriangleup V$, as a function of
  the peculiar velocity modulus. Solid lines show the results for the
  CDM particles at redshifts $z=0$ (red), $z=1$ (green) and $z=3$ (blue).
   Thick long-dashed lines represent the distribution of the peculiar velocities of the
  neutrino particles whereas thin short-dashed lines correspond to the
  unperturbed neutrino thermal velocity distribution at each
  redshift (Fermi-Dirac distribution).}
\label{PDF_velocities}
\end{figure}

We now test the validity of the above simple arguments using N-body
simulations. For a given simulation snapshot, we compute the modulus
of the peculiar velocity of all, CDM and neutrino, particles in
the box. We then calculate the fraction of particles, of each type,
whose peculiar velocity modulus lies between $V_{\rm{pec}}$ and
$V_{\rm{pec}}+\triangle V_{\rm{pec}}$, per $\triangle V_{\rm{pec}}$.
We show the results in the Fig. \ref{PDF_velocities} for the
simulation L60 (left panel) and L45 (right panel). We do not find
significant differences when we use the lower neutrino resolution
simulations LL60 and LL45, except in the very low velocity tail
($V_{\rm{peculiar}}\lesssim10$ km/s) and thus the results we show in
Fig. \ref{PDF_velocities} are numerically converged. Solid lines show
the results for the CDM particles at $z=3$ (blue), $z=1$ (green) and
$z=0$ (red). On average, CDM particles increase the modulus of their
peculiar velocities as time passes and structure formation
progresses. The results for the neutrino particles are represented by
the thick long-dashed curves and their behavior is opposite to the CDM one: on
average, neutrinos decrease their peculiar velocities as times evolves
as a consequence of the universe expansion. In order to study the
impact on the neutrino velocity field of processes such as neutrino
clustering within CDM haloes or neutrino evacuation from cosmological
voids, we plot with thin short-dashed lines the unperturbed distribution for the
neutrino peculiar velocities obtained from Eq. \ref{eq1}:
\begin{equation}
\frac{d f}{d V^{\nu}_{\rm{ther}}}(z)=\left(\frac{m_\nu}{1.803~c~k_B~T_\nu(z)}\right)\frac{x^2}{e^x+1}~,
\label{NU_pdf_vel}
\end{equation}
where $x=p_\nu/(k_BT_\nu(z))$ and $f(V,z)dV$ is the fraction of
neutrinos with thermal velocities between $V$ and $V+dV$. We have used
the approximation $p_\nu\cong m_\nu V_{\rm{ther}}^{\nu}/c$, with
$V_{\rm{ther}}^{\nu}$ the modulus of the neutrino thermal velocity,
which is very accurate once neutrinos are non-relativistic. At
redshifts $z>3$, the fully non-linear distribution of the modulus of
the neutrino peculiar velocities is very well described by the
unperturbed cosmic neutrino distribution of Eq. \ref{NU_pdf_vel}.
However, at lower redshift, the peculiar velocities of some neutrinos
are small enough to allow them to cluster within the gravitational
potential wells of CDM halos or to evacuate the interior of
cosmological voids. For that reason, the fraction of neutrinos with
low velocities will be smaller in the fully non-linear regime than in
the linear regime, since it is likely that those neutrinos will gain
gravitational energy.  Since the thermal velocities of relic neutrinos
increase as their masses drop, the deviation of the neutrino velocity
distribution from the unperturbed velocity distribution of Eq. \ref{NU_pdf_vel}
 distribution becomes smaller for smaller neutrino masses.  We find a constant 
 suppression in the fraction of neutrino particles with low velocities with respect
to the unperturbed distribution (Eq. \ref{NU_pdf_vel}). At redshift
zero, the proportion of neutrinos with peculiar velocities smaller
than 100 km/s is a factor $\cong2.08,~1.72,~1.33,$ and $1.15$ smaller
than the Fermi-Dirac distribution for neutrinos with $\Sigma_i
m_{\nu_i}=0.60,~0.45,~0.30,$ and $0.15$ eV, respectively.  We also
find that the maximum of the actual distribution is shifted with
respect to the unperturbed distribution. At $z=0$, the peak of the
actual velocity distribution is located at
$V\cong640,~805,~1145,~2220$ km/s for neutrinos with $\Sigma_i
m_{\nu_i}=0.60,~0.45,~0.30,~0.15$ eV, respectively, whereas the peak
in the unperturbed distribution is placed at
$V\cong560,~745,~1120,~2235$ km/s, respectively. The results for the
simulation with $\Sigma_i m_{\nu_i}=0.05$ eV are very close to those
from the simulation with $\Sigma_i m_{\nu_i}=0.15$ eV. We shall see on
Sec. \ref{Convergence_tests} that the neutrino dynamics (of the
massive species) in both simulations are basically the same. The high
velocity tail is, in all cases, very well reproduced by the
unperturbed neutrino velocity distribution.

\begin{figure}
\begin{center}
\includegraphics[width=0.495\textwidth]{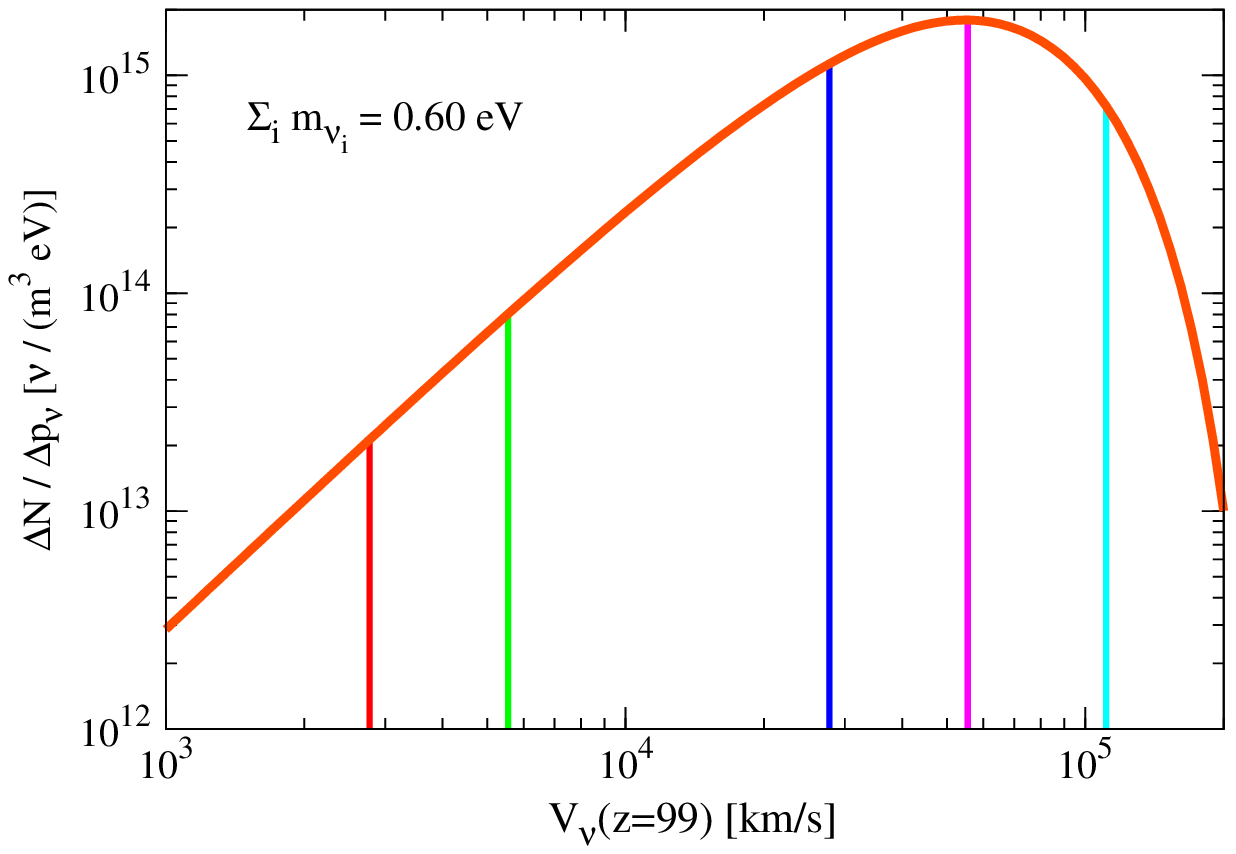}
\includegraphics[width=0.495\textwidth]{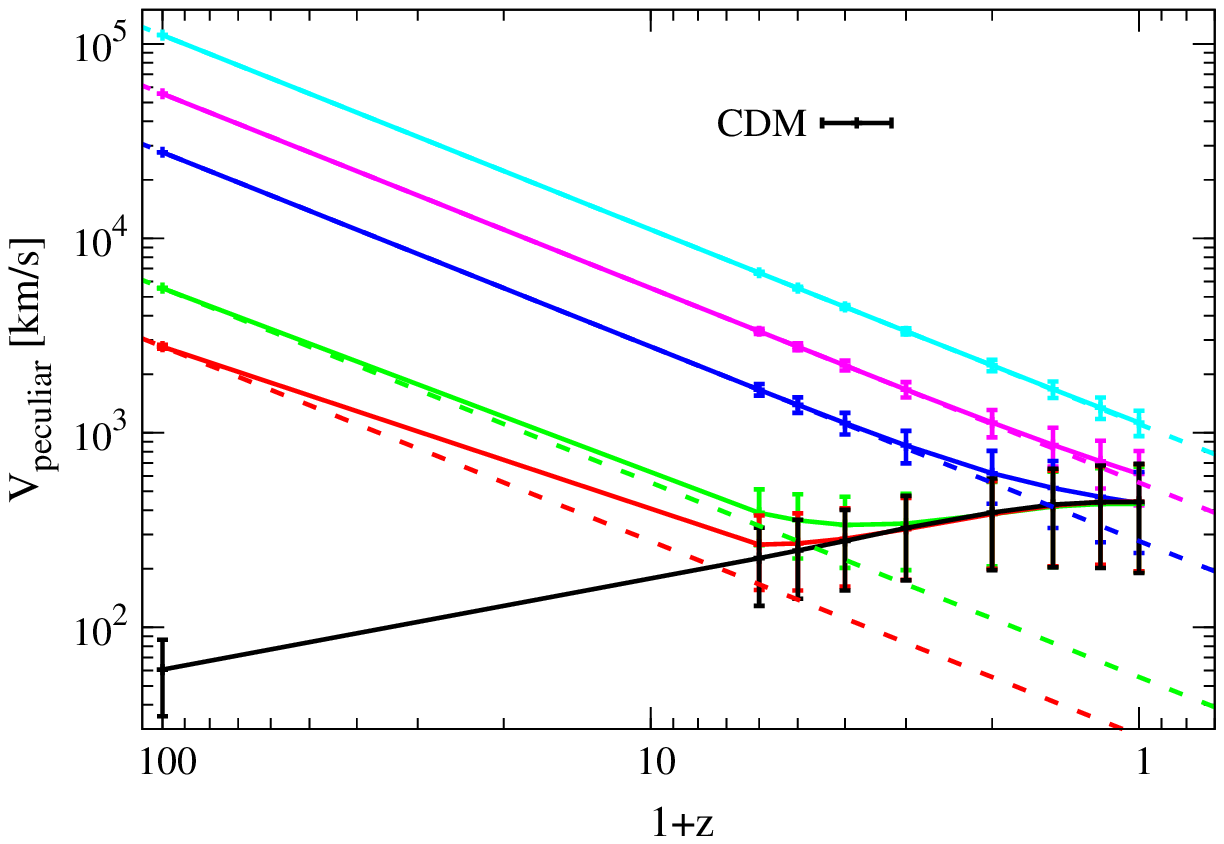}
\end{center}
\caption{For neutrinos with $\Sigma_i m_{\nu_i}=0.60$ eV, we show in
  the left panel with a solid orange line, the number density of
  neutrinos per unit of momentum as a function of the neutrino
  thermal velocity at $z=99$. As a function of redshift, the points and error bars 
 in the right panel represent the mean and the velocity dispersion of the
  neutrino particles, whose thermal velocities at $z=99$ lie within the narrow 
  velocity intervals that we show with vertical lines on the left panel (see text for 
  details). The dashed lines in the right panel show the unperturbed evolution of 
  the neutrino particles whose thermal velocities at $z=99$ lie within the 
  different velocity intervals of the left panel. The black line represents the mean 
  velocity and the velocity dispersion of a random but representative set of CDM 
  particles as a function of the redshift.}
\label{Velocity_evolution}
\end{figure}

We have investigated in detail the behavior of the neutrinos with low
thermal velocities over redshift. In the left panel of
Fig. \ref{Velocity_evolution} we plot with a solid orange line the
number of neutrinos with $\Sigma_i m_{\nu_i}=0.60$ eV and with
momentum between $p$ and $p+\triangle p$, per $\triangle p$ and per cubic meter, at
$z=99$, as a function of the neutrino thermal velocity modulus. At
this redshift, we take different velocity intervals that we show with
vertical lines in the same panel. For a particular velocity interval,
at $z=99$, we find all the neutrino particles whose thermal velocities
lie within it and store their IDs (integer numbers used to identify
particles along the simulation). We then use those IDs to find the
neutrino particles whose peculiar velocities belonged to a particular
velocity interval at $z=99$, and compute their peculiar velocities at
a posterior time. We perform this procedure at different redshifts and
in the right panel of Fig.  \ref{Velocity_evolution}, we plot the mean
and the velocity dispersion of the neutrinos whose thermal velocities
lie within the different velocity intervals at $z=99$. The color of
the points and curves are used to distinguish the different velocity
intervals at $z=99$. The dashed lines show the unperturbed time
evolution of the neutrinos, whose velocities drop as $1/(1+z)$. We
have also taken a representative set of CDM particles at $z=99$ and
followed their evolution along time (black solid line in the right
panel of Fig.  \ref{Velocity_evolution}). By representative we mean
that this set is small, in comparison with the total number of CDM
particles in the simulation, but large enough to make sure that the
quantities we compute are converged. Whereas neutrinos with initial
large thermal velocities follow very well the unperturbed evolution
(see lines in purple and cyan), neutrinos with lower velocities behave
in a different way. Once neutrinos are cold enough, their mean
velocity, velocity dispersion and velocity evolution become the same
as the this from the CDM. The redshift at which neutrinos catch up the
behavior of the CDM depends on their initial momentum as can be seen
in the right panel of Fig. \ref{Velocity_evolution}. We find that
neutrinos with velocities $\sim1/20$ and $\sim1/10$ of the mean
neutrino thermal velocity at $z=99$, start behaving as CDM at
redshifts $z\sim 3$ and $z\sim 2$, respectively. We therefore conclude
that the deviations in the neutrino peculiar velocity distribution
from the unperturbed distribution of Eq.~\ref{eq1} are mainly driven
by neutrinos with low momentum, which at some redshift start behaving
similarly to CDM.

\subsection{The halo mass function}
\label{HMF_sec}
We study the effects of massive neutrinos on the halo mass function
(MF) and compare with the ST (Sheth-Tormen) \cite{PS,ST}
prediction. This has been already presented in
\cite{Brandbyge_haloes,Marulli}. In \cite{Brandbyge_haloes} authors
limited their study at $z=0$. In \cite{Marulli}, the impact of the
neutrino masses on the halo mass function was studied at different
redshifts but the simulations were performed with the grid
implementation (see \cite{Viel_2010}), which can not capture fully the
non-linear neutrino regime at small scales. Here we aim at improving
over previous studies by computing the impact of neutrino masses on
the halo mass function at different redshifts, using the particle
implementation for the neutrino particles and for a very wide range of
halo masses.

\begin{figure}
\begin{center}
\includegraphics[width=0.49\textwidth]{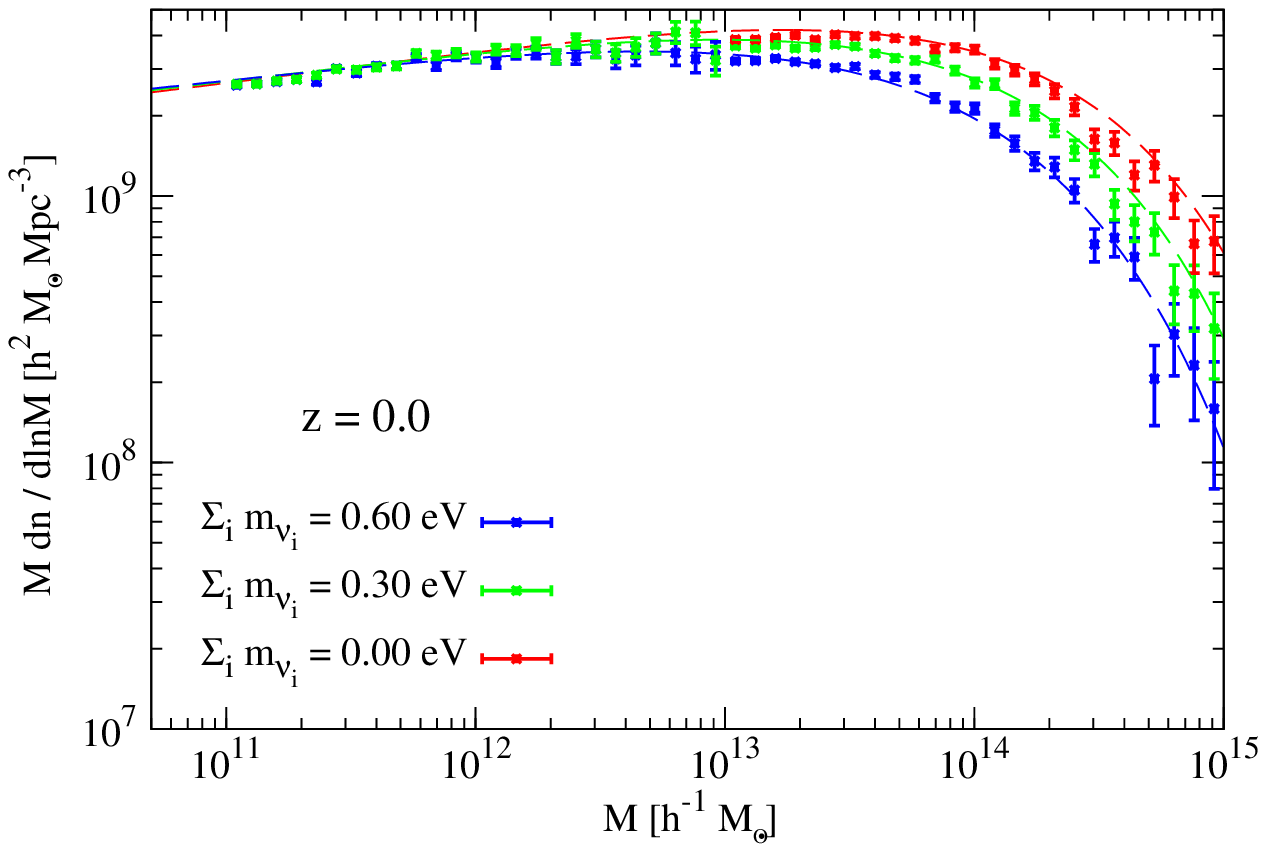}
\includegraphics[width=0.49\textwidth]{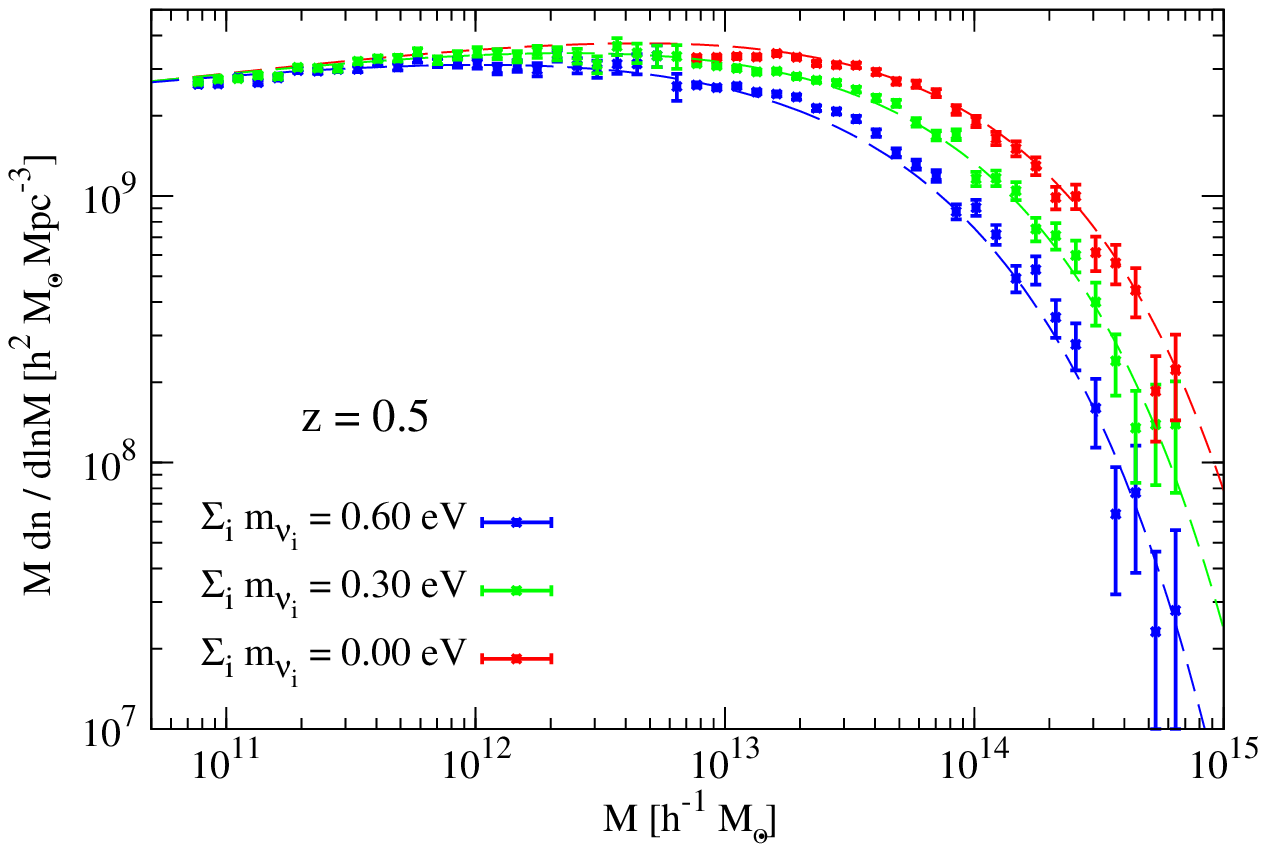}\\
\vspace{0.1cm}
\includegraphics[width=0.49\textwidth]{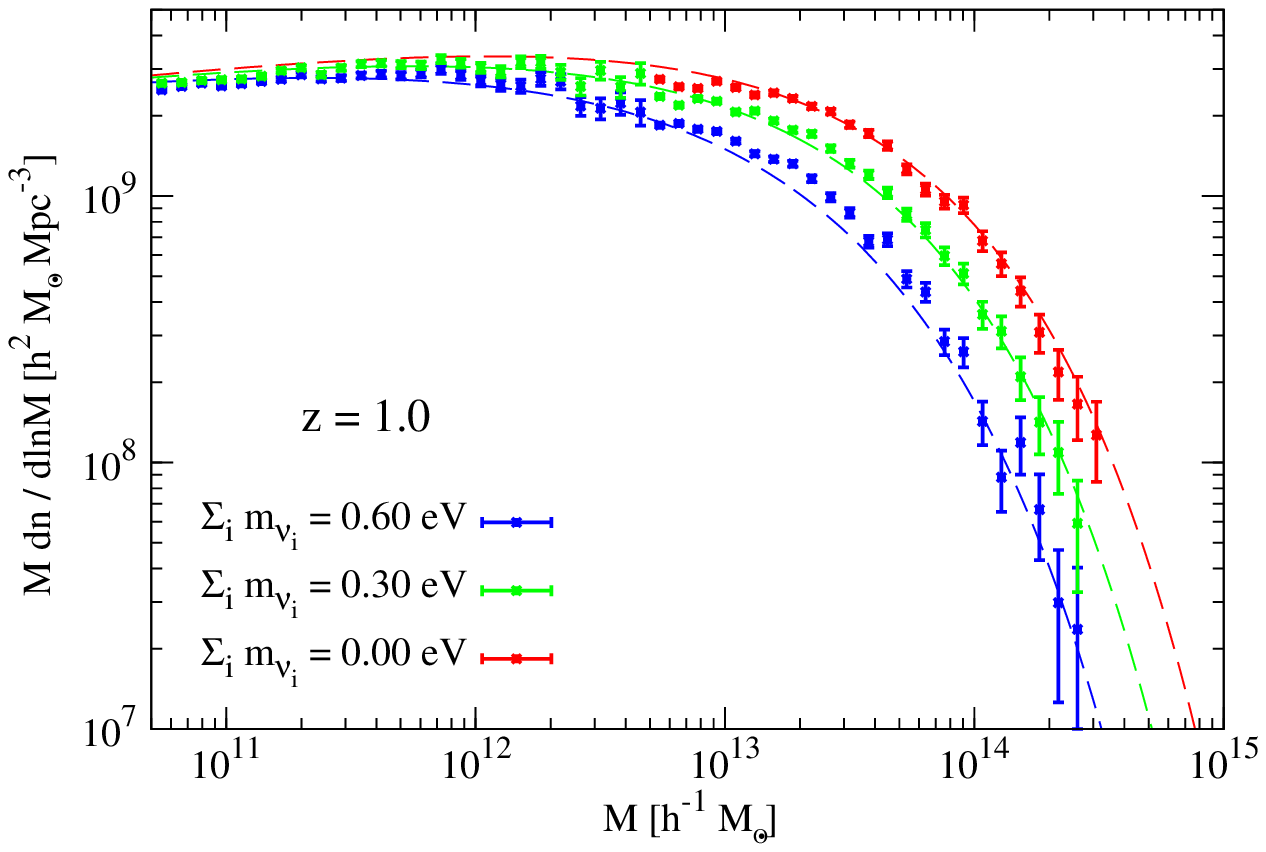}
\includegraphics[width=0.49\textwidth]{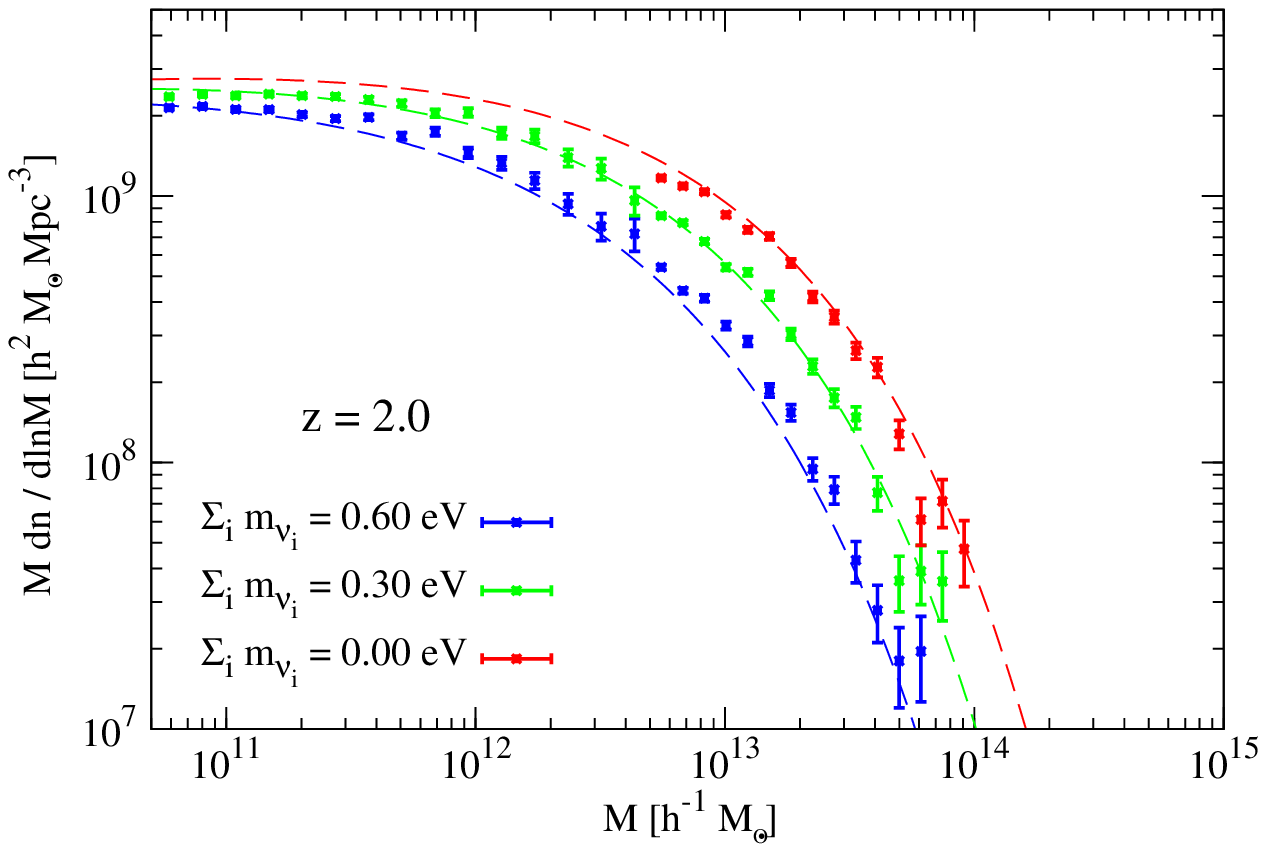}\\
\end{center}
\caption{Halo mass function for different cosmologies with different
  $\Omega_\nu$ satisfying
  $\Omega_{\rm{CDM}}+\Omega_{b}+\Omega_\nu=0.2708$. In all the models
  the amplitude of the power spectra is fixed to the same value on
  large scales (thereby the simulations have different $\sigma_8$
  values). The mass functions are shown for $z=0,0.5,1,2$ in the top
  left, top right, bottom left and bottom right panels, respectively,
  for two different total neutrino masses of 0.30 eV (green points) and
  0.60 eV (blue points). The massless neutrino case is shown with red
  points. The theoretical modified Sheth-Tormen predictions are shown
  as dashed curves.}
\label{HMF}
\end{figure}

From the N-body simulations we identify the CDM haloes and extract
their properties by applying the SUBFIND algorithm \cite{Subfind}.  A
CDM halo corresponds to a group identified by SUBFIND, where its
virial radius is defined as $M=\frac{4\pi}{3}\triangle_{\rm
  vir}\overline{\rho}_{\rm c}(z)R_{\rm vir}^3$ with $\triangle_{\rm
  vir}$ being the value of the mean overdensity at the time of
virialisation as predicted by the top-hat collapse model
\cite{BryanNorman}, $\triangle_{\rm vir}=18\pi^2+82x-39x^2$, where
$x=-\Omega_\Lambda/(\Omega_{\rm m}(1+z)^3+\Omega_\Lambda)$. The linear matter power spectrum, that impacts on the ST formula through
$\sigma(M)$, is computed using CAMB \cite{CAMB} taking the same
cosmological parameters of the N-body simulations.

The MFs from N-body simulations and the ST fits are plotted on
Fig. \ref{HMF} and the error bars correspond to the statistical
Poisson noise. The mass functions are extracted from both the low
resolution simulations (L60, L30 and L0) and the high resolution
simulations (S60 and S30) at different redshifts: $z=0, 0.5, 1$ and
$2$. The agreement between N-body simulations and the ST formula is
reasonably good as already found by
\cite{Brandbyge_haloes,Marulli}. However, for neutrinos with $\Sigma_i
m_{\nu_i}=0.60$ eV, we find a significantly larger halo abundance in
the N-body simulations than in the ST prediction. Those deviations are
larger for haloes of $\sim10^{13}$ $h^{-1}$M$_\odot$, independently of
redshift. The ST profiles are obtained by using the value of
$\Omega_{\rm {CDM}}+\Omega_{\rm{b}}$ as $\Omega_{\rm{M}}$ instead of
$\Omega_{\rm {CDM}}+\Omega_{\rm{b}}+\Omega_{\nu}$.  This point was 
discussed in detail in \cite{Brandbyge_haloes}. The reason why $\Omega_{\rm 
m}$ should be computed without including the neutrino component is because 
the clustering of neutrinos within CDM haloes is very small, as we will
see in the following Section. Thus, it is a very good approximation to
assume that neutrinos, at least for the masses considered here, do not
participate to the clustering process and neglect their contribution
to the $\Omega_{\rm M}$ value.

\subsection{The neutrino halo}
\label{neutrino_halo_sec}

In this Section we compute the neutrino density profiles around 
CDM haloes. Further, we provide the reader with a fitting function
that reproduces the neutrino profiles with high accuracy over a wide
range of radii.

The average velocity dispersion of CDM haloes is well described by the
formula \cite{Evrard}:
\begin{equation}
\sigma_{DM}(M,z)=\sigma_{DM,15}\left[\frac{h(z)M_{200}}{10^{15}\rm{M}_\odot}\right]^\alpha~,
\end{equation}
where $h(z)=H(z)/(100~\rm{km~s^{-1}Mpc^{-1}})$, M$_{200}$ is the mass
within the virial radius (in this case defined as the radius at
which the mean density is 200 times larger than the critical density),
$\sigma_{DM,15}$ and $\alpha$ are constants with values $\sim1080$ km/s 
and $\sim0.336$ respectively. Thus, at redshift $z=0$,
CDM haloes have typical velocity dispersions ranging from $\sim100$
km/s, for halo masses of $10^{12}~h^{-1}\rm{M}_\odot$, to $\sim 1000$
km/s for halo masses of $\sim10^{15}~h^{-1}\rm{M}_\odot$. For a fixed
CDM halo mass, the velocity dispersion grows with redshift as
$h(z)^\alpha$.

On the other hand, the mean thermal velocity of neutrinos
is equal to $160(1+z)($eV/$m_\nu$) km/s, while their velocity
dispersion reads:
\begin{equation}
\sigma_\nu=c\left(\frac{\rm{eV}}{m_\nu}\right)\sqrt{\frac{\int_0^\infty n_\nu(p,z)(p-\overline{p}_\nu(z)^2)dp}{\int_0^\infty n_\nu(p,z)dp}}~,
\end{equation}
which implies $\sigma_\nu\sim
87(1+z)\left(\frac{\rm{eV}}{m_\nu}\right)~\rm{km/s}$. Therefore, we
would expect relic neutrinos to cluster within the gravitational
potential wells of CDM haloes, at least for the most massive
neutrinos. This clustering will be larger for higher neutrino masses,
and we would expect that it starts at low redshift, since at high
redshift the low velocity dispersion of CDM haloes and the large
neutrino thermal velocities will prevent it.

The clustering of neutrinos within the gravitational potential wells
of CDM haloes has already been studied using semi-analytic models
\cite{Ma,Wong,Paco,Ichiki} and N-body simulations \cite{Brandbyge_haloes}. In
\cite{Brandbyge_haloes}, it was demonstrated that the agreement
between both methods is fairly good. The differences found were likely
due to the simplified assumptions used in \cite{Wong} to compute the
neutrinos clustering, e.g. the fact that the gravitational potential
wells of CDM were assumed not to change with time. A more realistic
calculation can be found in \cite{Paco}. Here we compute the neutrino
density profiles within CDM haloes using a different set of N-body
simulations, with slightly different values of the cosmological
parameters and with the improved code {\sc GADGET}-3.

\begin{figure}
\centering
\includegraphics[width=0.8\textwidth]{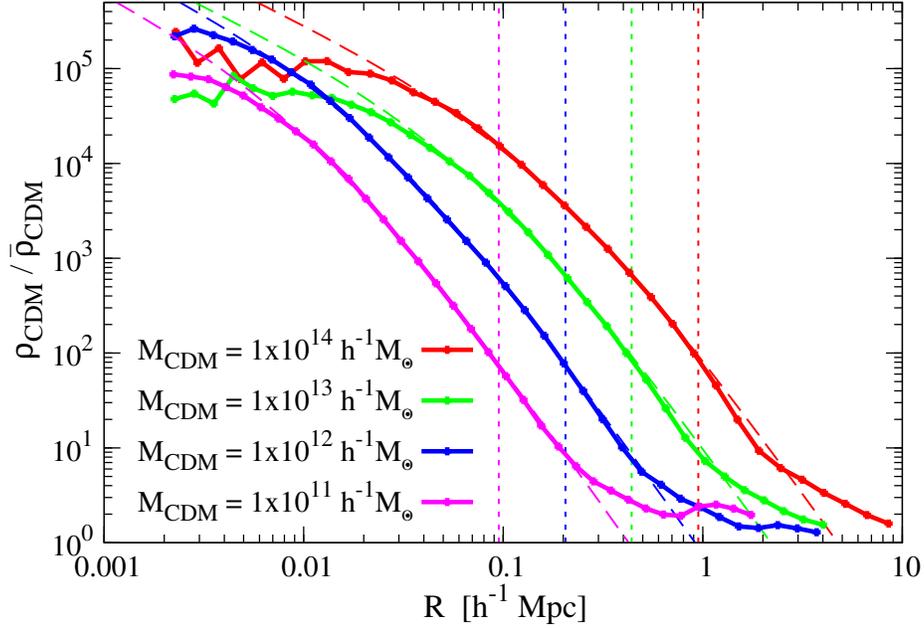}
\caption{Average CDM density profiles at $z=0$. The points show the
  mean density profile for the CDM component as a function of radius
  for four different CDM halo masses at redshift zero:
  $1\times10^{14}$ (red), $1\times10^{13}$ (green), $1\times10^{12}$
  (blue) and $1\times10^{11}$ (purple) $h^{-1}$M$_\odot$ (for each
  mass we select all CDM haloes within a mass bin of width $5\%$  ). 
  The vertical lines correspond to the values
  of the virial radius for the different haloes, whereas the dashed
  lines show the NFW profile that best fits each profile.}
\label{profiles_CDM}
\end{figure}

The CDM haloes are identified from the N-body simulations as groups by
the algorithm SUBFIND (see Sec. \ref{HMF_sec} for further details). We
focus our study on the clustering of neutrinos within isolated CDM
haloes. Our choice of focusing on isolated CDM haloes is deliberate:
the clustering of neutrinos is very sensitive to the CDM distribution
on large scales, therefore, by focusing on isolated CDM haloes we will
obtain results that will have less dispersion than in the general
case. In Appendix \ref{Appendix_A}, we show how the neutrino profiles
change when we compute the clustering of neutrinos within non-isolated
CDM haloes.

We define a CDM halo as isolated if no more massive CDM haloes are
situated at a distance less than 10 times its virial radius If the former
condition is not satisfied, then the CDM halo is non-isolated. We study the 
clustering of relic neutrinos within
CDM haloes of different masses at $z=0$: $10^{11}$, $10^{12}$,
$10^{13}$ and $10^{14}$ $h^{-1}$M$_\odot$. For a given mass of the
host CDM halo, we create a halo catalog consisting of all haloes whose
virial masses differ from this value by less than $5\%$. This
tolerance is chosen in order to increase the number of haloes over
which we compute quantities, i.e. improve the statistical significance
of the results, while keeping it low enough to avoid selecting haloes
with very different properties (such as the value of their virial
radius). For a given mass of the host CDM halo we
compute the density profile for the CDM component for all of its halo
members, at $z=0$, and in Fig. \ref{profiles_CDM} we show the mean density 
profiles for the different halo
catalogs. The profiles shown in the figure have been extracted from
the simulation L60, i.e. for the cosmology with $\Sigma_i
m_{\nu_i}=0.60$ eV. We find that the average CDM density profiles are
very well fitted by the Navarro-Frenk-White (NFW) profile \cite{NFW}
$\rho_{NFW}(r)=\frac{\rho_s}{(r/r_s)(1+r/r_s)^2}$,
up to the virial radius. We have also computed the average CDM density
profiles for the cosmologies with $\Sigma_i m_{\nu_i}=0.45$, 0.30,
0.15 and 0.00 eV and it turns out that, in all cases, the mean density
profiles are almost identical. However, we find that the halo
concentration, $c=R_{vir}/r_s$, slightly decreases with the neutrino
masses. This behavior was also found by \cite{Brandbyge_haloes} and, as
they discussed, it is likely due to the fact that massive haloes form
at later times in cosmologies with massive neutrinos, as we have seen
from the halo mass function in Sec. \ref{HMF_sec}. 

\begin{figure}
\begin{center}
\includegraphics[width=0.73\textwidth]{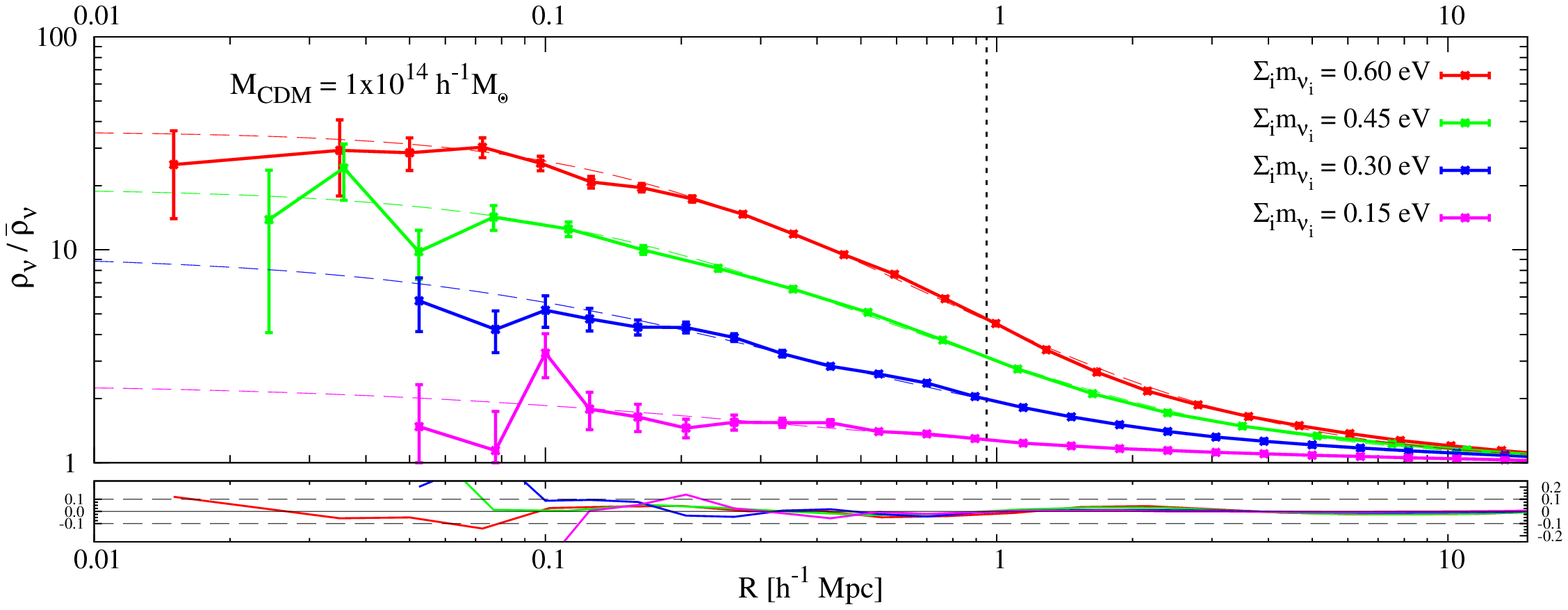}\\
\includegraphics[width=0.73\textwidth]{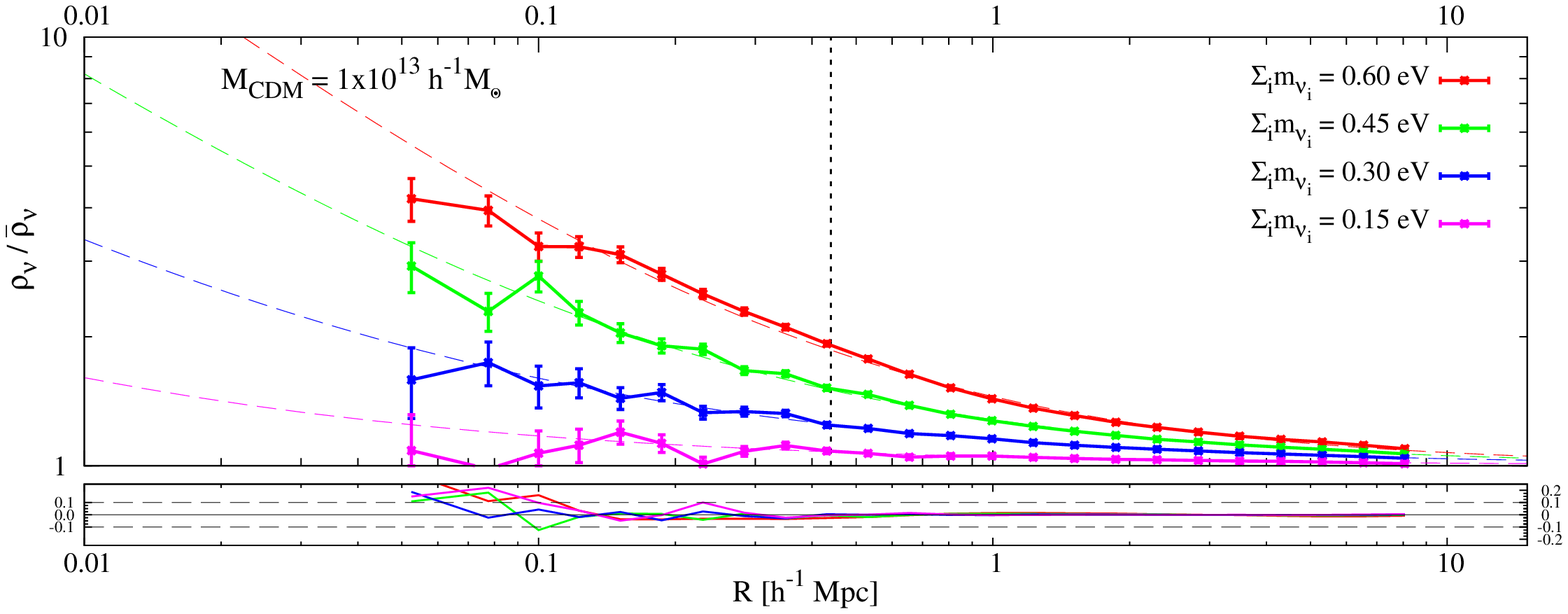}\\
\includegraphics[width=0.73\textwidth]{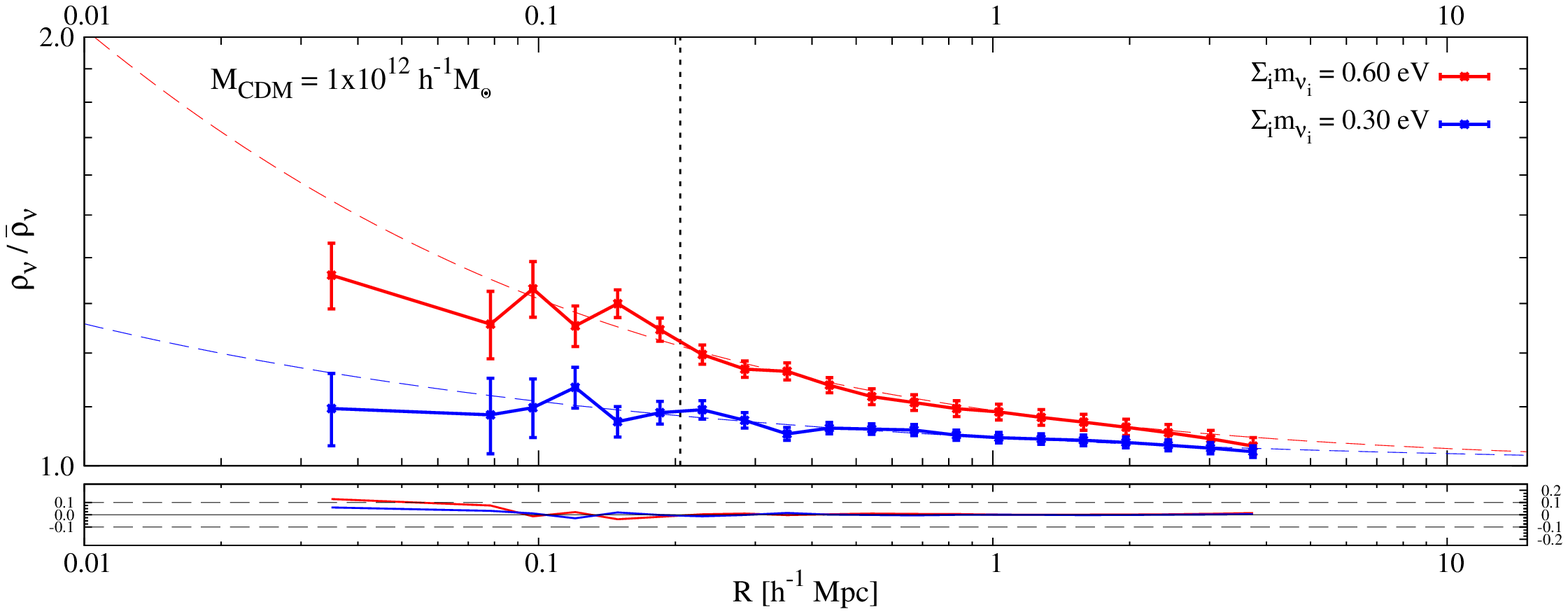}\\
\includegraphics[width=0.73\textwidth]{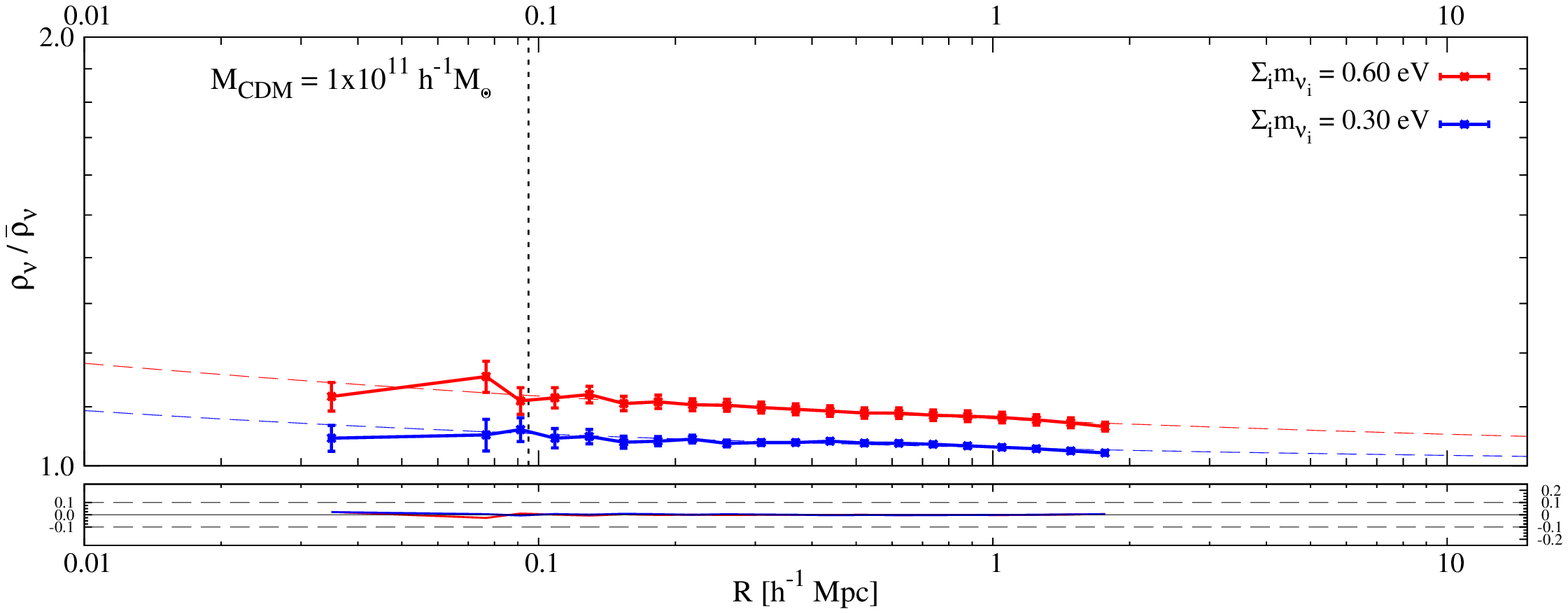}\\
\end{center}
\caption{Mean neutrino overdensity profiles for different neutrino
  masses and for different masses of their host CDM haloes. Each panel
  shows the average neutrino density profiles at $z=0$, normalized to
  the mean neutrino density, that arise due to the clustering of
  neutrinos within CDM haloes of masses $1\times10^{14}$
  $h^{-1}M_\odot$ (first panel), $1\times10^{13}$ $h^{-1}M_\odot$
  (second panel), $5\times10^{12}$ $h^{-1}M_\odot$ (third panel),
  $1\times10^{11}$ $h^{-1}M_\odot$ (fourth panel) at $z=0$. The error
  bars show the dispersion of the mean overdensity profile.  The
  profiles are computed for cosmologies with $\sum_i m_{\nu_i}=$0.60
  eV (red), 0.45 eV (green), 0.30 eV (blue) and 0.15 eV
  (purple). Dashed lines represent the profiles of Eq. \ref{VBV} that
  best fit the computed average neutrino overdensity profiles. On the
  bottom of each panel we plot the relative difference between the
  profiles and the fitting formula. The value of the virial radius for the
  different host CDM halo masses is shown in each panel with a vertical
  line.}
\label{NU_profiles_m}
\end{figure}

We repeat the same procedure for the neutrino component and in
Fig. \ref{NU_profiles_m} we show the average neutrino density profiles
at $z=0$ normalized to the neutrino background density. The error bars
represent the dispersion in the average neutrino overdensity profile,
that we compute in the following way: for a given halo catalog, we
compute the mean overdensity profile and the dispersion around it. We
then assume that, for a given radius bin, the overdensity values are
distributed following a Gaussian distribution, and thus, the
dispersion in the mean overdensity profile will be given by
$\sigma^{\rm{profile}}_\nu(r)/\sqrt{N}$, where N is the number of
halos in the catalog and $\sigma^{\rm{profile}}_\nu(r)$ is the
dispersion around the mean overdensity profile. The neutrino density
profiles are computed for each host CDM halo mass extracted from the
simulations L60, L45, L30 and L15 for four different neutrino masses:
$\Sigma_i m_{\nu_i}=0.60, 0.45, 0.30$ and $0.15$ eV, respectively. We
shall see in Sec. \ref{Convergence_tests} that the overdensity
profiles for a cosmology with $\Sigma_i m_{\nu_i}=0.05$ eV would be
equivalent to those obtained from L15. The noisy behavior that arises
at small scales is due to the finite number of particles and to the
finite number of CDM haloes. Therefore, on small scales, the
dispersion in the average neutrino density profiles is relatively
large, pointing out our resolution limits.

For a given sum of the neutrino masses, the clustering of relic
neutrinos increases with the masses of their host CDM haloes, since
more massive haloes have larger and deeper gravitational potential
wells than less massive ones. For a fixed mass of the host CDM halo,
the higher the neutrino masses the larger the neutrino clustering will
be: this is because the proportion of neutrinos with low or moderate
peculiar velocities, the ones subjected to clustering, increases with
the neutrino masses. In fact, due to their large thermal velocities,
neutrinos are not able to cluster on small scales: thus, the neutrino
profiles are not cuspy but they exhibit a relatively large core.

For Milky way size haloes ($\sim10^{12}$ $h^{-1}$M$_\odot$) our numerical resolution do not allow us to explore the inner regions of the neutrino density profile. We therefore conclude that the relic neutrino overdensity at the solar radius has to be larger than $\sim40\%$, with respect to this of the background, for neutrinos with $\Sigma_i m_{\nu_i}=0.60$ eV and above $\sim10\%$ for neutrinos with $\Sigma_i m_{\nu_i}=0.30$ eV.

Although the values of the cosmological parameters in
\cite{Brandbyge_haloes} are different to ours, we find that our
results are in good agreement with their results. We also find a
reasonable agreement with the neutrino overdensity profiles computed
through the N-one-body method described in \cite{Wong,Paco}.

\begin{figure}
\begin{center}
\includegraphics[width=0.49\textwidth]{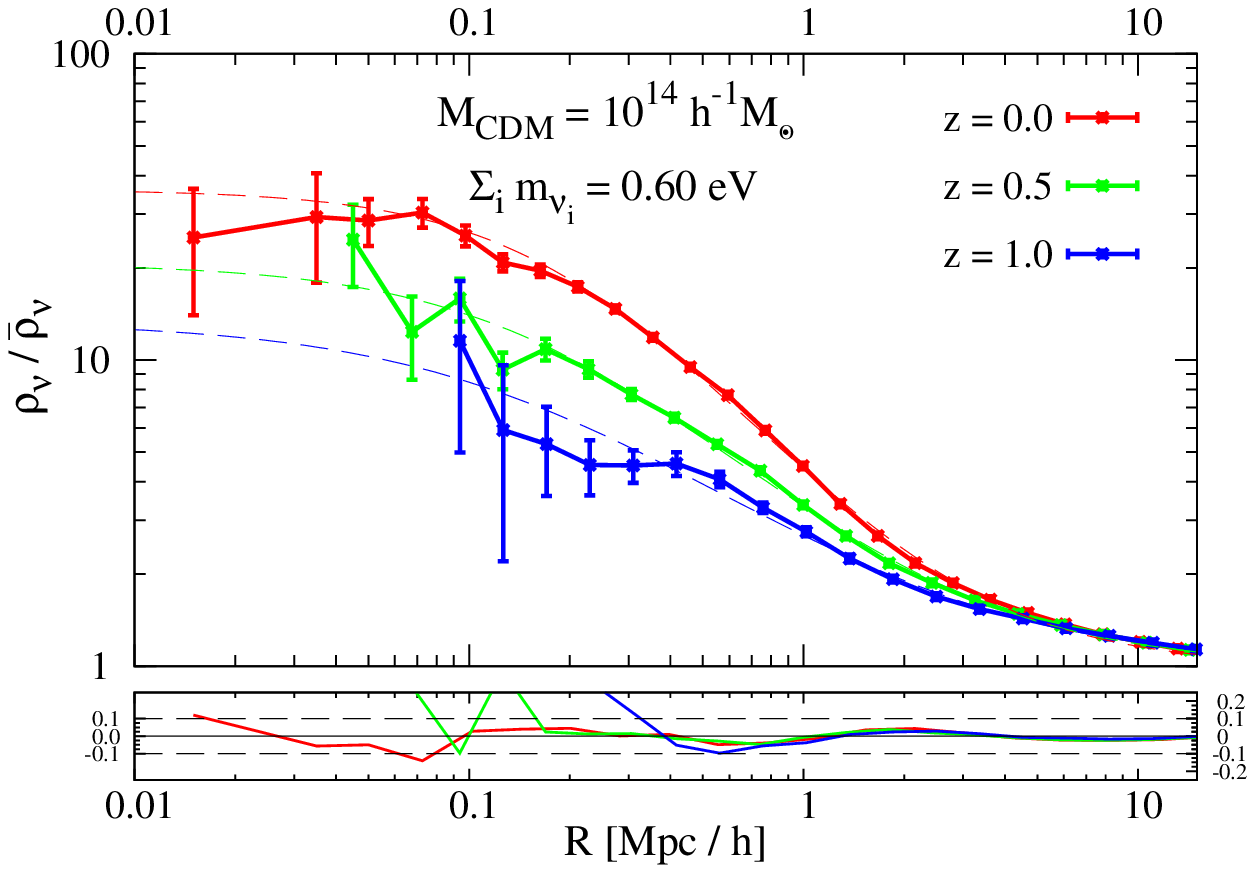}
\includegraphics[width=0.49\textwidth]{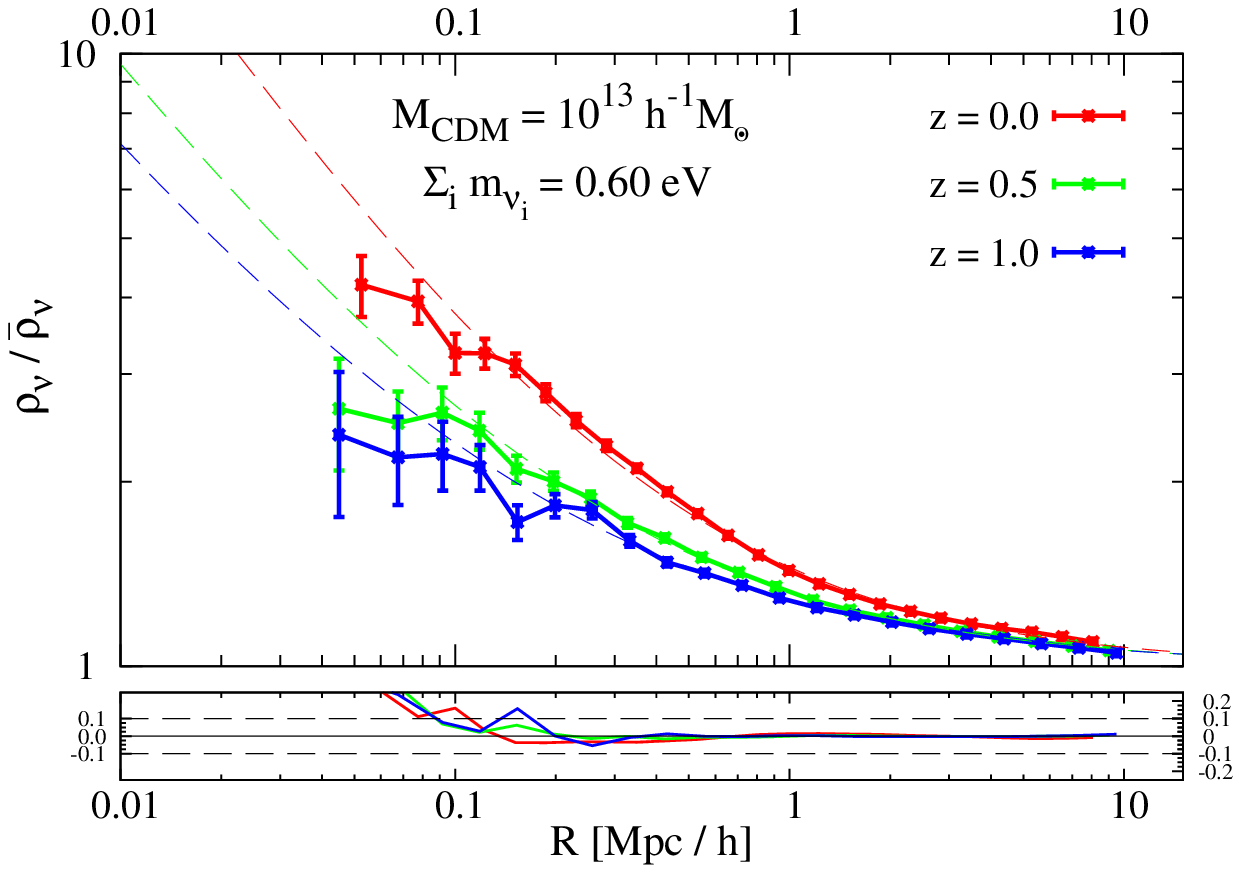}\\
\end{center}
\caption{Dependence of the neutrino clustering with redshift for a
  fixed mass of the CDM halo. We plot the neutrino overdensity
  profiles at redshifts $z=0$ (red), $z=0.5$ (green) and $z=1$ (blue)
  within CDM haloes that have masses equal to $10^{14}$
  $h^{-1}$M$_\odot$ at $z=0$ (red), $z=0.5$ (green) and $z=1$
  (blue). The right panel shows the same quantities but for CDM haloes
  of masses $10^{13}$ $h^{-1}$M$_\odot$. In both panels the density
  profiles refer to the $\Sigma_i m_{\nu_i}=0.60$ eV case.}
\label{NU_profiles_z}
\end{figure}

We also investigate the dependence of the neutrino density profiles
with redshift for a fixed mass of the host CDM halo. That is done by
computing the neutrino density profiles within CDM haloes that have
the same virial mass at different redshifts. The results are shown in
Fig. \ref{NU_profiles_z} for two different masses of the host CDM halo
and for the $\Sigma_i m_{\nu_i}=0.60$ eV case. In the left panel, the
red line represents the average neutrino density profile at $z=0$,
computed for host CDM haloes that have a virial mass equal to
$10^{14}$ $h^{-1}$M$_\odot$ at $z=0$; the green (blue) line
corresponds to the average neutrino density profile at $z=0.5~(z=1)$,
computed within CDM haloes that have a virial mass equal to $10^{14}$
$h^{-1}$M$_\odot$ at $z=0.5~(z=1)$; the $x-$axis represents the
distance to the halo center in comoving units. The right panel shows
the same for host CDM haloes with masses equal to $10^{13}$
$h^{-1}$M$_\odot$.
 
Although CDM haloes have the same masses, the thermal velocities of
neutrinos are larger at high redshift\footnote{Note also that the
  CDM halo velocity dispersion grows with redshift ($\propto
  h(z)^\alpha$) more slowly than the mean and velocity dispersion of
  relic neutrinos ($\propto (1+z)$)}. As a consequence, the clustering
of neutrinos becomes smaller as redshift increases.

\subsubsection{Fitting function}

We find that the average neutrino overdensity profiles are well
described by the following equation:
\begin{equation}
\delta_\nu(r)=\frac{\rho_\nu(r)-\overline{\rho}_\nu}{\overline{\rho}_\nu}=\frac{\rho_c}{1+(r/r_c)^\alpha}
\label{VBV}
\end{equation}
over a wide range of radii. The physical meaning of the parameters in
the profile \ref{VBV} is very simple: $r_{\rm c}$ and $\rho_{\rm c}$
represent the length and the overdensity of the core in the
overdensity profile of the neutrino halo while $\alpha$ is a parameter
that controls how fast the overdensity profile falls on large
radii. However, for CDM halo masses below $\sim10^{13.5}$
$h^{-1}$M$_\odot$, the resolution in our N-body simulations is not
large enough to properly resolve the core in the neutrino density
profiles {\footnote{We note that a core is always expected
    because of the Tremaine-Gunn bound \cite{TG_bound}}. This gives
rise to a degeneracy between the parameters $\rho_{\rm c}$ and $r_{\rm
  c}$. We find that a simple profile of the form:

\begin{figure}
\begin{center}
\includegraphics[width=0.90\textwidth]{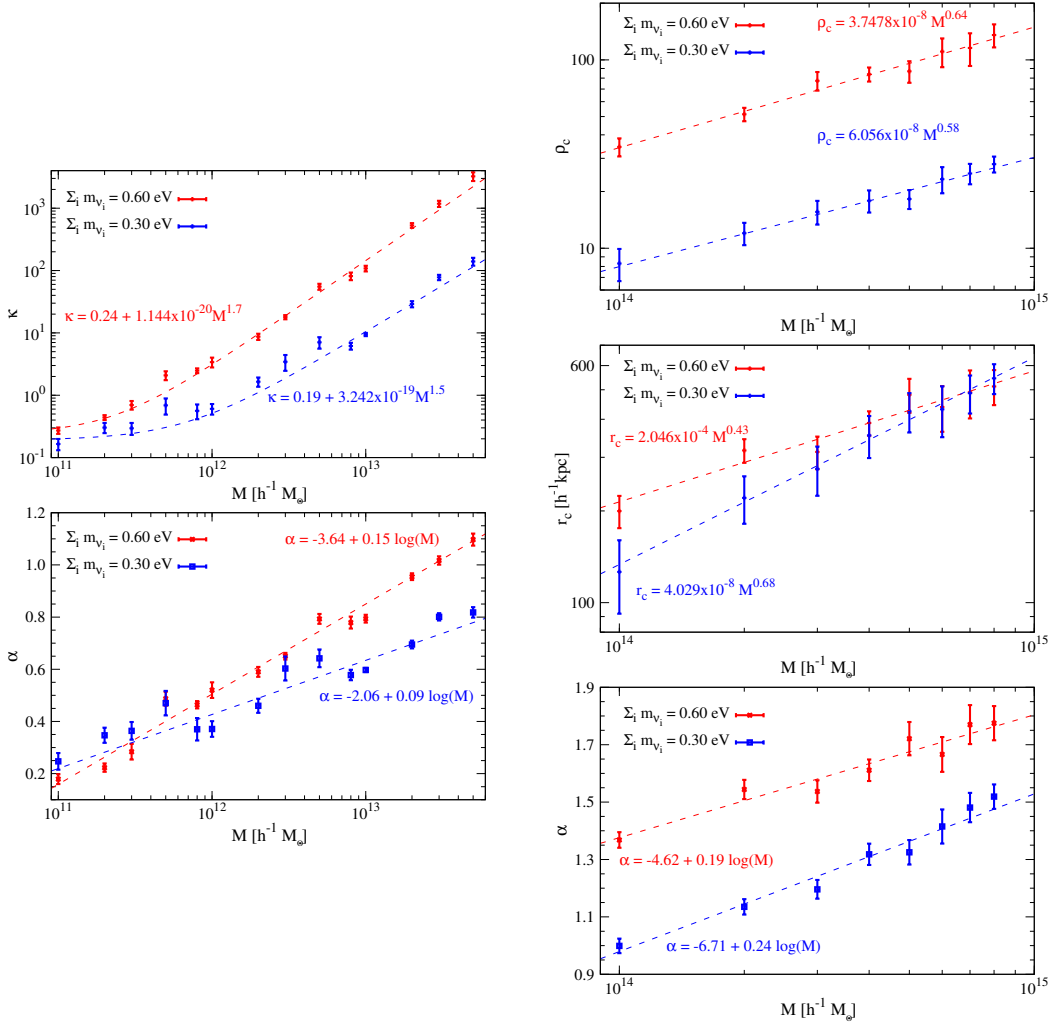}\\
\end{center}
\caption{Derived parameters from the fitting function used in
    order to investigate the neutrino density profile. The panels on the right
  show the dependence of the parameters of the profile of Eq. \ref{VBV} with the 
neutrino masses and the masses of their host CDM halo. For lighter CDM halos, 
our N-body resolution do not allow us to resolve properly the parameters 
$\rho_{\rm c}$ and $r_{\rm c}$ (see text for details). The panels on the left
show the dependence of the parameters for the simpler profile of 
Eq.~\ref{VBV_simple} with the neutrino masses and CDM halo masses.}
\label{VBV_fig}
\end{figure}

\begin{equation}
\delta_\nu(r)=\kappa/r^\alpha~,
\label{VBV_simple}
\end{equation}
reproduces the outskirts of the computed neutrino density profiles very well. 
Note that on distances much larger than the core radius, $r\gg r_{\rm c}$, the 
profile \ref{VBV} reduces to \ref{VBV_simple}, with $\kappa=\rho_{\rm c} r_{\rm c}^\alpha$. 
For a given average neutrino overdensity profile, the values of the parameters 
$\overrightarrow{p}=(\rho_{\rm c}, r_{\rm c}, \alpha)$ ($\overrightarrow{p}=(\kappa, \alpha)$ 
for CDM halos masses below $\sim10^{13.5}$ $h^{-1}$M$_\odot$) are those that minimize the quantity:
\begin{equation}
\chi^2=\sum_i\left[\frac{\delta_\nu^i-\delta_v(r_i,\overrightarrow{p})}{\sigma^i_\nu}\right]^2~,
\end{equation}
where $\delta_\nu^i$ and $\sigma_\nu^i$ are the values of the average
neutrino overdensity and overdensity dispersion at radius $r_{\rm i}$,
respectively. The fitting formula, with the value of their parameters
extracted as above, are plotted as dashed lines in
Fig.~\ref{NU_profiles_m}. In the bottom part of the panels in
Fig.~\ref{NU_profiles_m} we plot the relative difference between the
fitting formula and the average density profiles. 

The outskirts of the neutrino haloes are very well reproduced by the
fitting formula for all CDM halo masses, although it works best for
lower CDM halo masses. At large radii, the fitting formula is more
accurate for small values of the neutrino masses. When using the
power-law profile (Eq.  \ref{VBV_simple}), important discrepancies
between the fitting formula and the neutrino density profiles take
place on small scales. This happens because the fitting formula of
Eq. \ref{VBV_simple} is cuspy, whereas the neutrino density profiles
must exhibit a core (see for example \cite{Wong}). We note that the
former formula will eventually violate the Tremaine-Gunn bound
\cite{TG_bound}, and for that reason, the extrapolated values of the
fitting formula should be taken with caution.

The values of the fit parameters depend on three quantities: the mass
of the host CDM halo, redshift and the masses of the neutrinos. We
investigate the dependence of the fitting formula parameters on
neutrino masses and on the mass of the host CDM halo. In
Fig.~\ref{VBV_fig} we plot the values of the fitting profile
parameters, extracted as explained above, as a function of the mass of
the CDM halo that hosts the neutrino halo for two different neutrino
masses. Both the CDM halo masses and the average neutrino profiles
from which the parameters are extracted are at $z=0$. The red points
represent the values of the parameters for neutrinos of masses
$\Sigma_i m_{\nu_i}=0.60$ eV, while the blue points are for neutrinos
of masses $\Sigma_i m_{\nu_i}=0.30$ eV. The error bars correspond to
the $1\sigma$ errors on the value of the parameters, while the dashed
lines represent a simple fitting formula that fits the values of the
parameters reasonable well (see appendix \ref{Appendix_B}).

We check whether the fitting formula of Eq.~\ref{VBV}
(\ref{VBV_simple} when the core is not properly resolved) can
reproduce the average density profiles of neutrino haloes at
$z>0$. The fitting profiles, whose parameter values are extracted as
above, are shown with dashed lines in Fig.~\ref{NU_profiles_z} for
different masses of the host CDM halo at different redshifts. The
relative difference between the density profile points and the fitting
formula is shown at the bottom of each panel on that figure: the
fitting profile reproduces very well the outskirts of the computed
profiles at all redshifts. The inner region is better described when
the neutrino clustering is large enough to properly resolve the core
in the density profile.

\subsubsection{Convergence tests}
\label{Convergence_tests}

We now investigate whether our results are numerically stable with
respect to: number of neutrino particles, size of the simulation box
and starting redshift of the simulation. In the left panel of
Fig.~\ref{NU_profile_convergence}, we show the results of computing
the mean neutrino density profiles at $z=0$, within isolated CDM
haloes of masses $10^{14}$ $h^{-1}M_\odot$ at redshift $z=0$, for
different neutrino masses, by using several simulations with the same
box size but different number of neutrino particles. In particular, we
compare the results from the simulations L60 and L45 to those obtained
from LL60 and L45. We find that by increasing the number of neutrino
particles by a factor eight, keeping fixed the size of the simulation
box, the density profiles vary less than $10\%$ for radii larger than
$\sim 200$ $h^{-1}$kpc. We therefore conclude that our results are
already converged for those radii even for the low resolution
simulations. At smaller radii, it turns out that the results are more
stable the more massive the neutrinos are. This is a consequence of
the discrete sampling of the neutrino phase-space: the lower the
neutrino masses the higher the neutrino phase-space distribution has
to be sampled to resolve the small scale features in the neutrino
density profiles.

In the left panel of Fig. \ref{NU_profile_convergence} we also show
the neutrino overdensity profiles extracted from the simulations L15
and LL5. Whereas the simulation L15 contains three degenerate species
of neutrinos having each of them a mass equal to 0.05 eV, the
simulation LL5 consists of two massless neutrino species and one
massive species with a total mass equal to 0.05 eV. Although the
matter power spectrum is different in both simulations (see table
\ref{tab_sims}), the neutrino momentum distribution and the CDM halo
velocity dispersion are basically identical. For that reason, we find
that the results are the same in both cases, being converged at the
$10\%$ level for radii larger than 200 $h^{-1}$kpc. We emphasize that this happens because we are
considering the neutrino overdensity, a quantity which is insensitive
to the number of degenerate species while the unnormalized neutrino
density profile is a factor three larger in L15 with respect to LL5.

\begin{figure}
\centering
\includegraphics[width=0.495\textwidth]{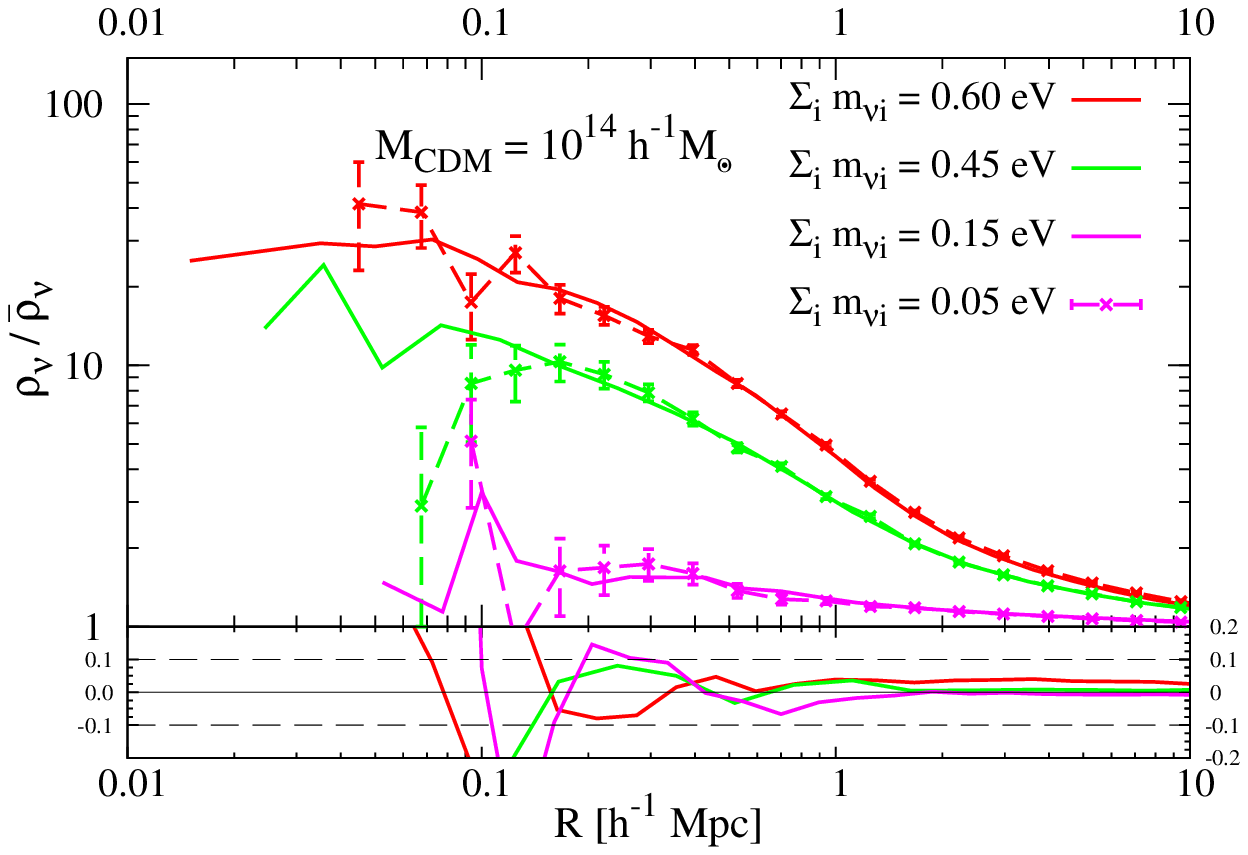}
\includegraphics[width=0.495\textwidth]{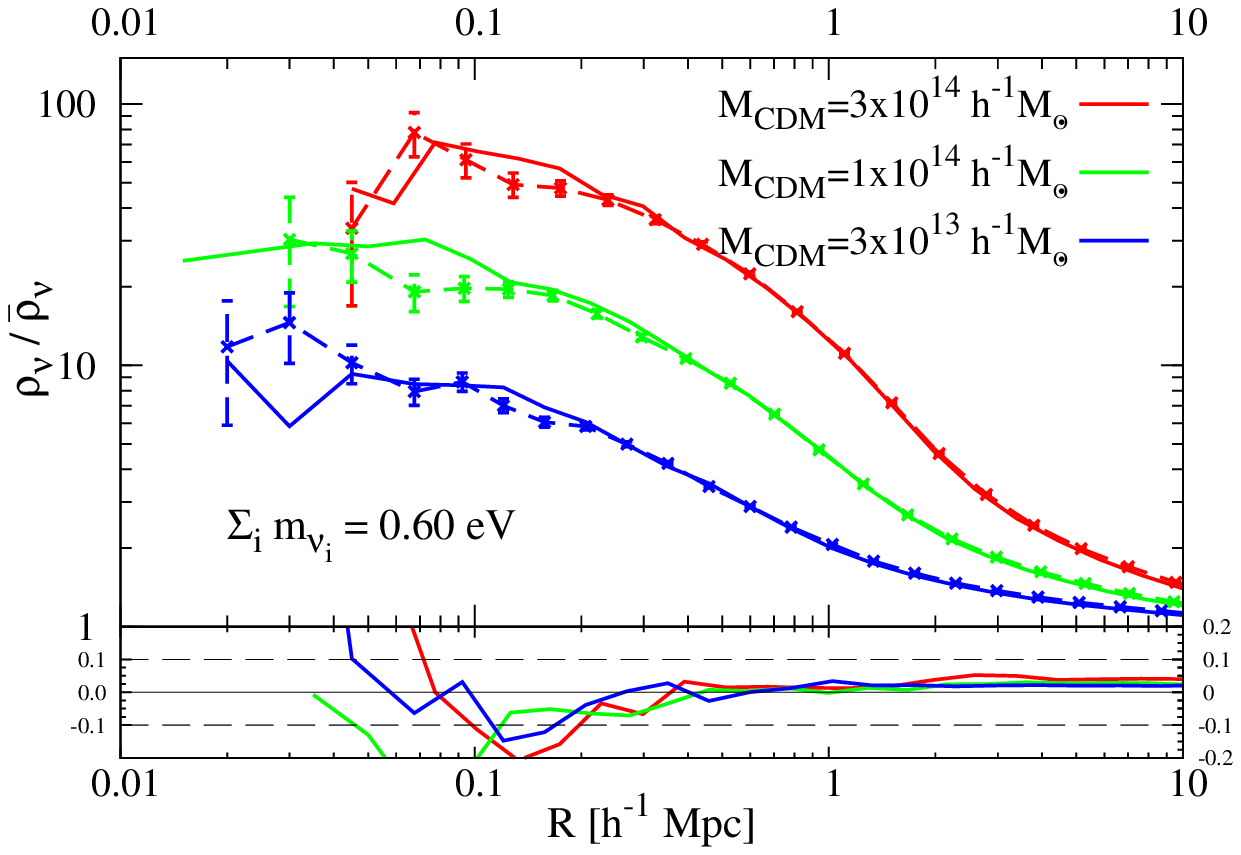}
\caption{\textit{Left panel}: Dependence of the neutrino overdensity
  profiles on the number of neutrino particles for a box of fixed
  size. The figure shows the mean neutrino density profiles for
  isolated CDM haloes of masses $10^{14}$ $h^{-1}M_\odot$, at $z=0$,
  extracted from the simulations L60, L45, L15 (solid lines) and LL60,
  LL45, LL5 (dashed lines). In the bottom panel, we plot the relative
  difference between the low-resolution and the high-resolution
  profiles. \textit{Right panel:} dependence of the neutrino
  overdensity profiles with the simulation box size and starting
  redshift. The solid lines show the mean neutrino density profile, at
  $z=0$, for different masses of their isolated host CDM haloes at
  $z=0$: $3\times10^{14}$ (red), $1\times10^{14}$ (green) and
  $3\times10^{14}$ $h^{-1}M_\odot$ (blue) extracted from the
  simulation L60. The dashed lines represent the same quantities but
  as extracted from the simulation H60. In the bottom panel, we plot
  the relative difference between the profiles obtained from H60 and
  those from L60. The error bars show the dispersion of the average
  neutrino overdensity profile for the simulations with lower
  resolution.}
\label{NU_profile_convergence}
\end{figure}

We repeat this analysis at different redshifts. We find that for CDM
haloes of masses $10^{14}$ $h^{-1}M_\odot$, the relative differences
between the profiles computed from the low and from the high
resolution simulation remain below $10\%$ for radii larger than 400
$h^{-1}$kpc at redshift $z=0.5$, whereas at redshift $z=1$ differences
become larger than $10\%$ for radii smaller than 700 $h^{-1}$kpc. At
this redshift, the relative difference between the models with
$\Sigma_i m_{\nu_i}=0.05$ eV and $\Sigma_i m_{\nu_i}=0.15$ eV keeps
below $10\%$ for radii larger than 400 $h^{-1}$kpc. The reason why
differences become larger for larger neutrino masses is because at
this redshift there are few CDM haloes with masses equal to $10^{14}$
$h^{-1}M_\odot$, and therefore, the computed profiles are prone to the discreteness in the neutrino phase-space and to the cosmic variance.

We also study the dependence of the average neutrino density profiles
on the size of the cosmological box and on the starting redshift of
the simulation. We check this by running a N-body simulation with a
box size of 1000 $h^{-1}$Mpc, $512^3$ CDM particles and $1024^3$
neutrino particles. The starting redshift is $z=19$, in contrast with
our default choice of $z=99$. The simulation corresponds to the
cosmological model with $\Sigma_i m_{\nu_i}=0.60$ eV and is listed on
the table \ref{tab_sims} as H60. In the right panel of
Fig.~\ref{NU_profile_convergence} we show with dashed lines the
average neutrino density profiles at $z=0$, for different masses of
their host CDM haloes, extracted from the simulation H60. We compare
those density profiles with those obtained from the simulation L60
(solid lines) and find that both profiles differ by less than $10\%$
for radii larger than 200 $h^{-1}$kpc, for all CDM halo masses.  It is
worth noting that in the simulation H60, the resolution of the CDM and
neutrino particles are a factor eight below that of the simulation
L60. The fact that the profiles do not depend on the initial redshift
of the simulation should not be surprising since we have already seen
that neutrino clustering starts at very recent
times \footnote{However, the starting redshift of the simulation has
  to be high enough for 2LPT to properly describe the CDM evolution.}.

\begin{figure}
\centering
\includegraphics[width=0.495\textwidth]{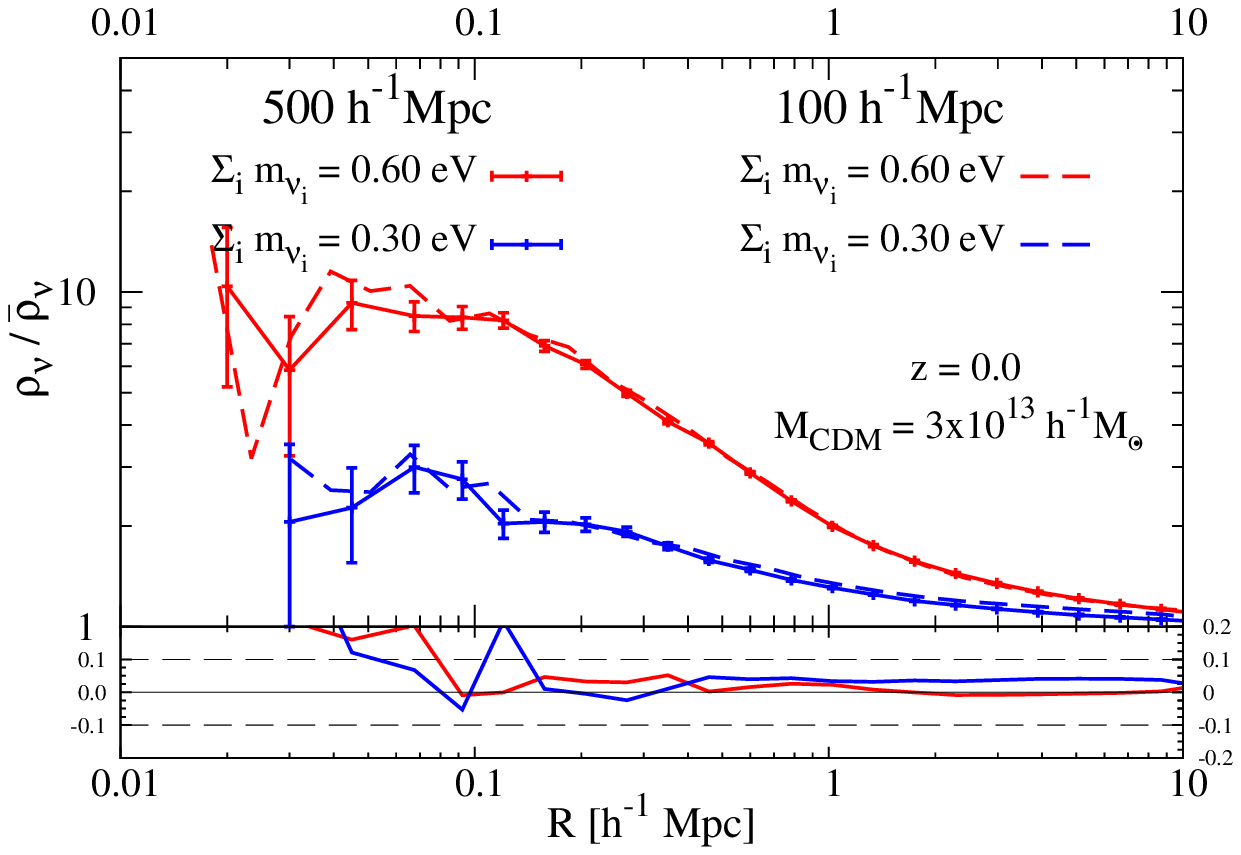}
\includegraphics[width=0.495\textwidth]{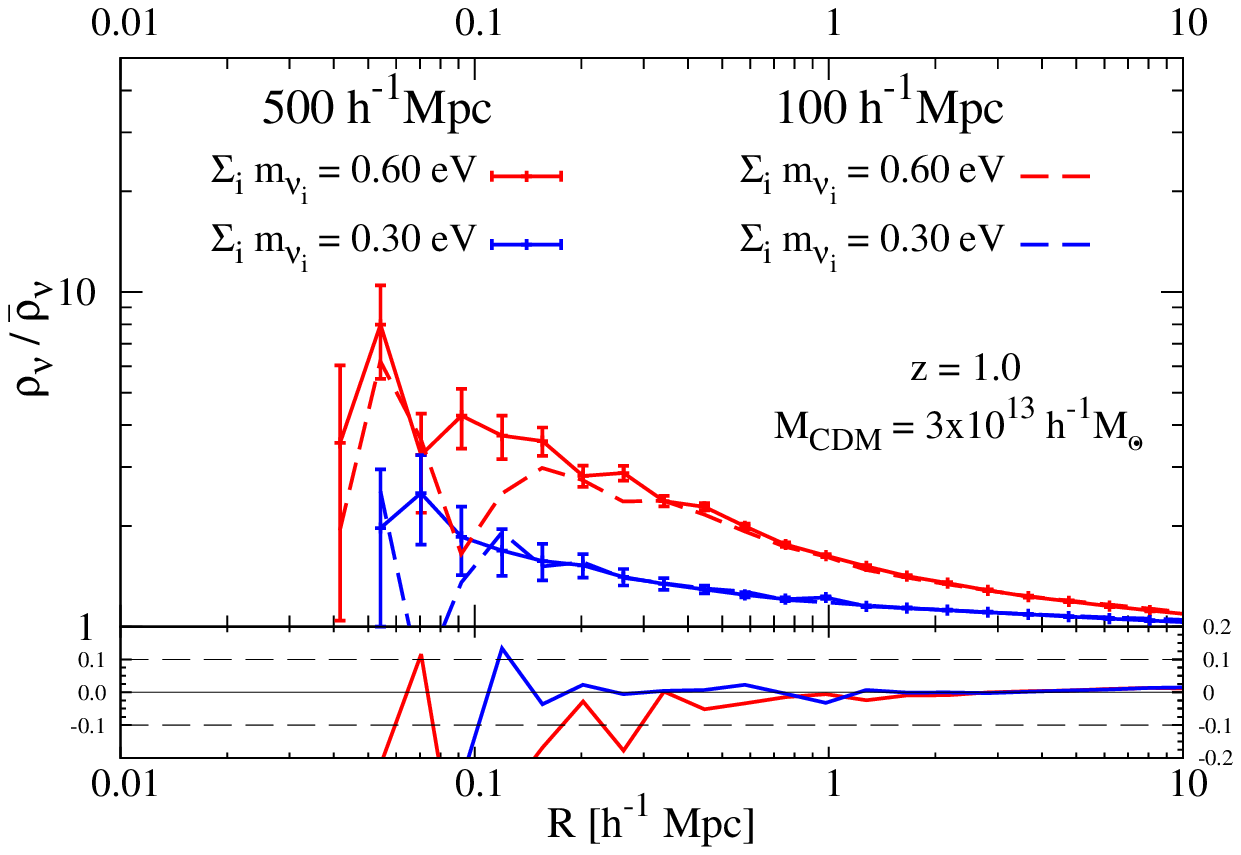}
\caption{Comparison between the results of the simulations L60 and 30
  (solid lines) and the results of the simulations S60 and S30 (dashed
  lines). For neutrinos with $\Sigma_i m_{\nu_i}=0.60$ eV (red) and
  $\Sigma_i m_{\nu_i}=0.30$ eV (blue) we plot the average neutrino
  density profiles, within CDM haloes of masses equal to
  $3\times10^{13}$ $h^{-1}$M$_\odot$ at $z=0$ (left) and at $z=1$
  (right), extracted from the the simulations L60, S60, L30 and
  S30. At the bottom we show the relative difference between the
  profiles. The error bars show the dispersion of the average neutrino
  overdensity profile for the simulations with lower resolution.}
\label{High_low}
\end{figure}

Finally, we compare the neutrino density profiles extracted from the
low resolution N-body simulations L60 and L30 to those obtained from
the high resolution simulations S60 and S30. Since the size of the box
in the simulations S60 and S30 is five times smaller than that of L60
and L30, the comparison can only be performed for CDM haloes of masses
$\sim 10^{13}$ $h^{-1}$M$_\odot$. In Fig.~\ref{High_low} we plot the
average neutrino density profiles, at redshifts $z=0,1$ within CDM
haloes of masses equal to $3\times10^{13}$ $h^{-1}$M$_\odot$ extracted
from the high and low resolution simulations. We find a very good
agreement between both simulations. For CDM haloes of masses
$(1-2)\times10^{13}$ $h^{-1}$M$_\odot$ we find some discrepancies in
the average neutrino density profiles, arising as a consequence of
slightly different average CDM density profiles. This is likely due to
the relatively low number of CDM haloes with those masses in the high
resolution simulations ($\sim 30$), which produces a slightly biased
result with respect to the average. We therefore conclude that the
results from the low resolution simulations are converged for CDM halo
masses larger than $1\times10^{13}$ $h^{-1}$M$_\odot$.

\subsection{Neutrino velocity distribution within CDM haloes}
In this Section we study the distribution of the neutrino peculiar
velocities within isolated CDM haloes. In Sec.~\ref{nl_velocity_field}
we have seen that the distribution of neutrino peculiar velocities
computed over the whole set of simulated boxes is reasonably well
described by the unperturbed neutrino momentum distribution of Eq.~
\ref{eq1}. However, at low redshift, the proportion of neutrinos with
low peculiar velocities is over-estimated by the unperturbed
distribution. As we showed in Fig.~\ref{Velocity_evolution}, this
happens because neutrino momenta can not keep decreasing their value
as $\propto 1/(1+z)$, since once neutrino velocities are low enough,
they will behave in the same way as CDM particles do.

By kinematical considerations, the proportion of neutrinos with low
velocities within CDM haloes has to be small. On the other hand, we
have seen that neutrino haloes are substantially more extended than
their CDM counterparts, since typical neutrino peculiar velocities are
larger than those of CDM.

\begin{figure}
\begin{center}
\includegraphics[width=0.49\textwidth]{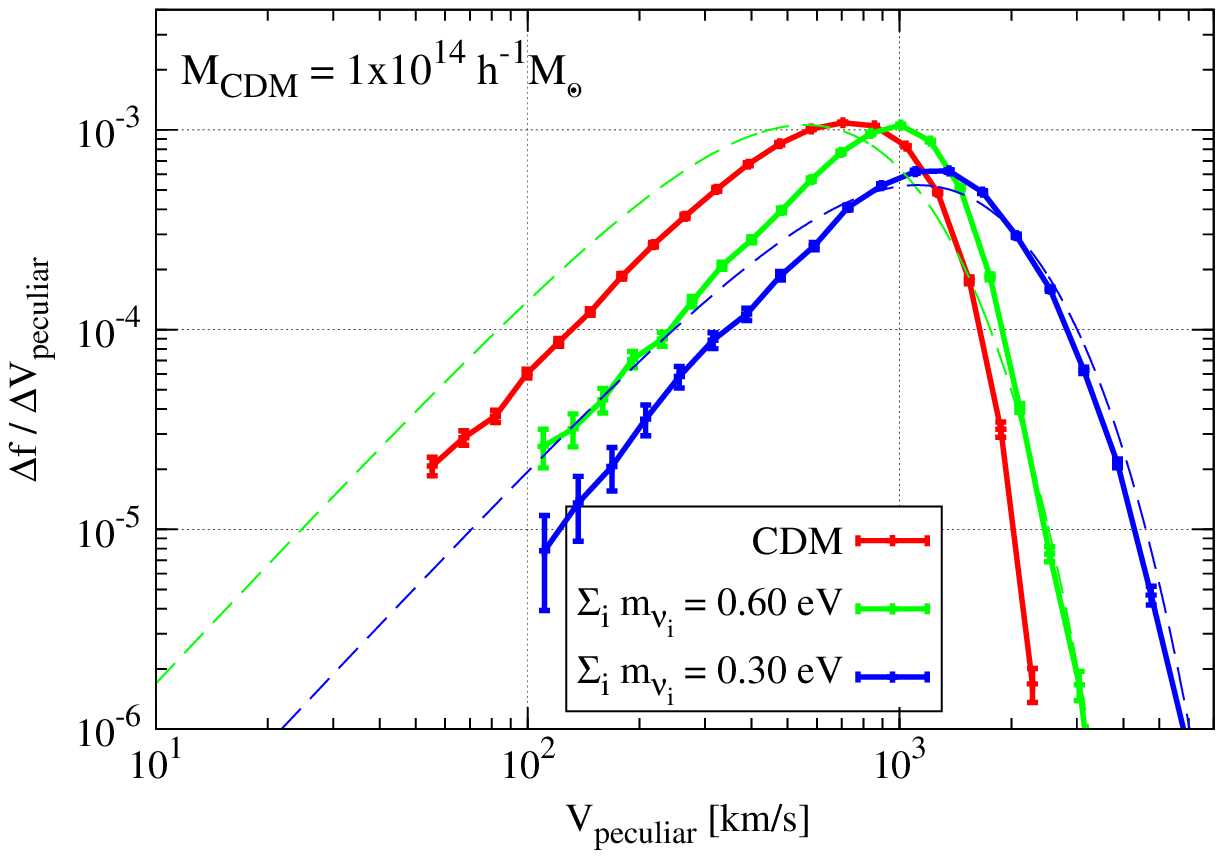}
\includegraphics[width=0.49\textwidth]{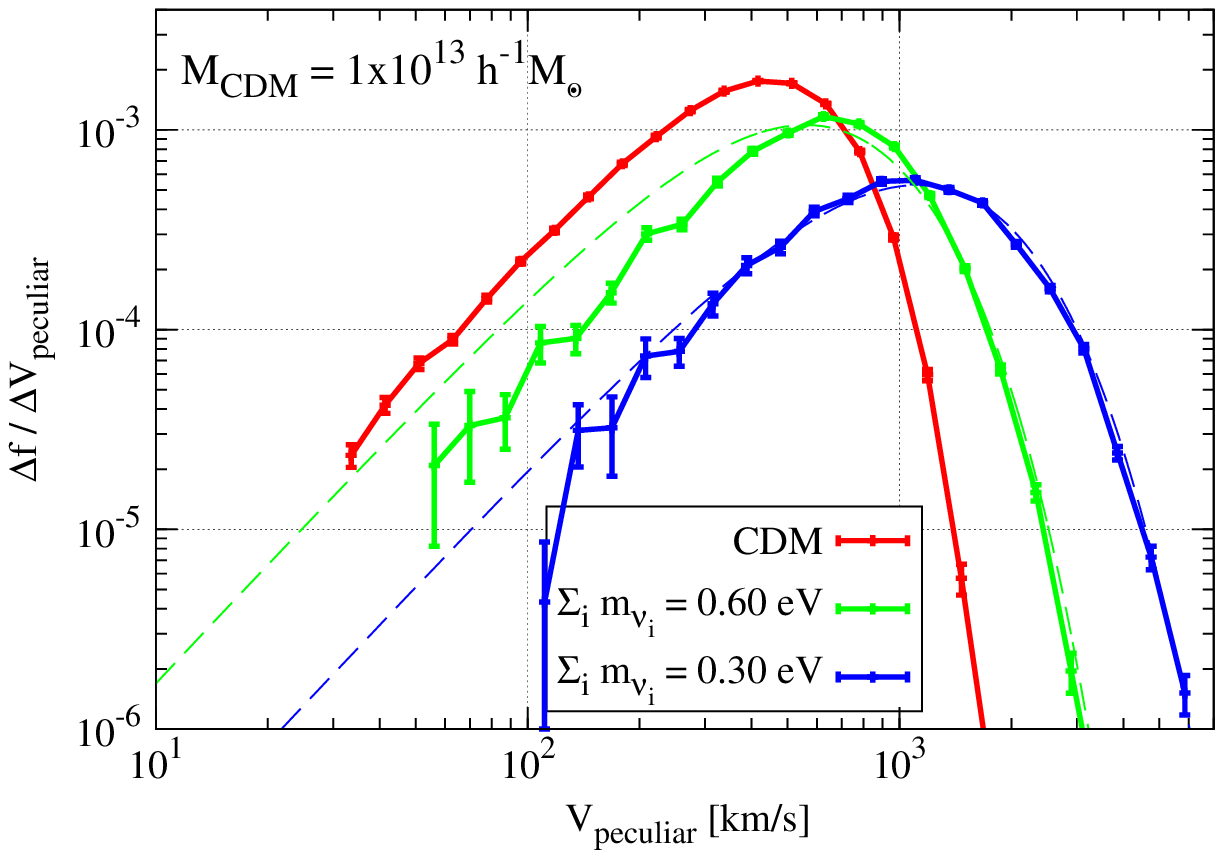}\\
\vspace{0.1cm}
\includegraphics[width=0.49\textwidth]{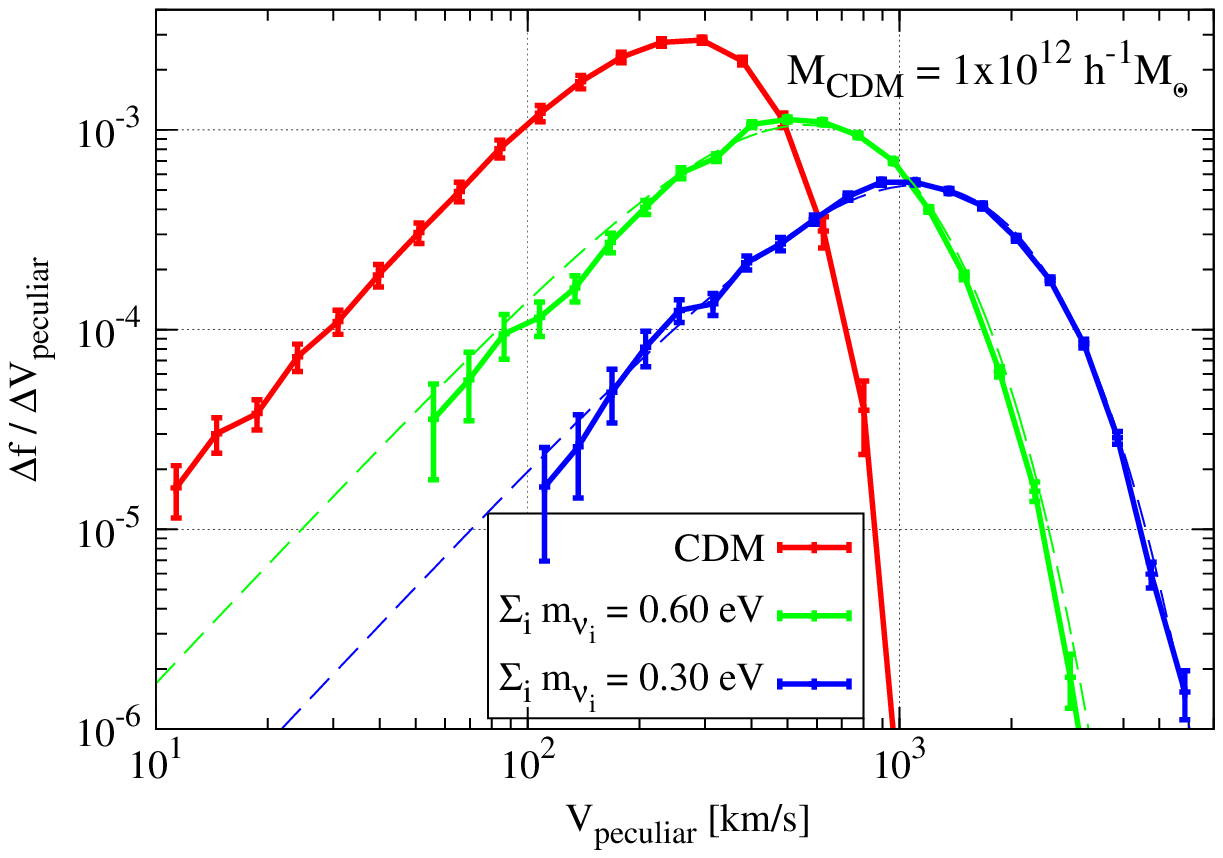}
\includegraphics[width=0.49\textwidth]{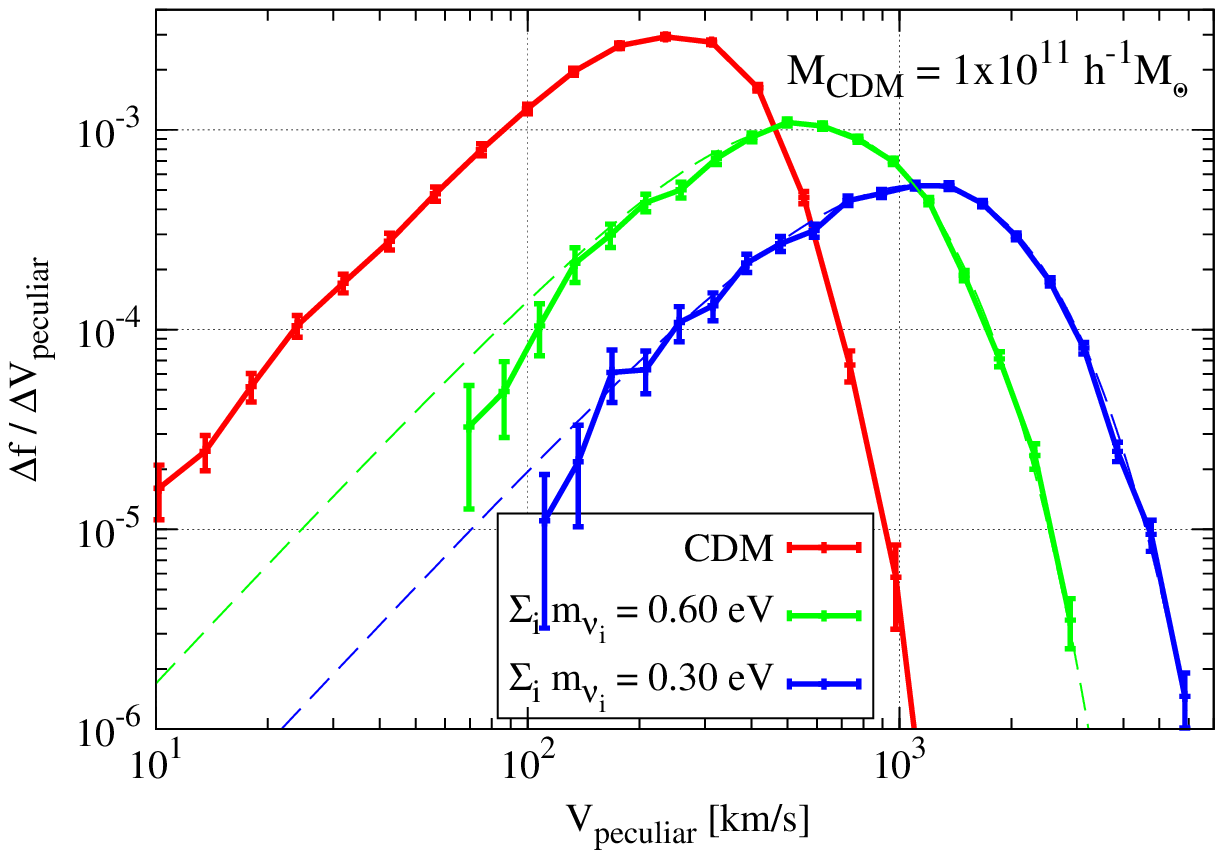}\\
\end{center}
\caption{Average CDM and neutrino velocity distribution within the
  virial radius of CDM haloes. We plot the proportion of particles
  with peculiar velocities between $V$ and $V+\Delta V$, per $\Delta
  V$, as a function of the modulus of the peculiar velocity. With
  solid lines we show the average velocity distribution within CDM
  haloes with masses equal to $1\times10^{14}$ (\textit{upper left}),
  $1\times10^{13}$ (\textit{upper right}), $1\times10^{12}$
  (\textit{bottom left}) and $1\times10^{11}$ (\textit{bottom right})
  $h^{-1}$M$_\odot$ at $z=0$, for the CDM particles (solid red line),
  neutrinos with $\Sigma_i m_{\nu_i}=0.60$ eV (green line) and
  neutrinos with $\Sigma_i m_{\nu_i}=0.30$ eV (blue line). The error
  bars represent the $1\sigma$ uncertainty in the average
  velocity distribution.  The dashed lines represent the unperturbed
  neutrino velocity distribution \ref{NU_pdf_vel}.}
\label{PDF_velocity_haloes}
\end{figure}

We compute the distribution of the neutrino peculiar velocities within
CDM haloes of different masses at $z=0$ by considering the
distribution of neutrino momenta within the CDM halo virial radius. As
in the case of the neutrino density profiles, for a given mass of the
host CDM halo, we create a halo catalog by selecting all isolated CDM
haloes whose masses differ of it by less than a $5\%$. For each CDM
halo belonging to a given catalog, we take all neutrino particles that
lie within the CDM halo virial radius and compute the fraction of
neutrinos in velocity bins. By fraction we mean the number of
particles within a velocity interval over the total number of
particles within the CDM halo virial radius. For a given sum of the
neutrino masses, the velocity intervals are chosen to be the same for
all haloes in a catalog. We repeat that procedure for all CDM haloes
in a given catalog and finally, for each velocity interval, we compute
the mean and the dispersion for all the obtained values. The results
are shown in Fig.~\ref{PDF_velocity_haloes} for four different masses
of the host CDM halo: $1\times10^{11}$, $1\times10^{12}$,
$1\times10^{13}$ and $1\times10^{14}$ $h^{-1}$M$_\odot$, and for two
neutrino masses, $\Sigma_i m_{\nu_i}=0.30$ and 0.60 eV. We also
compute the velocity distribution for the CDM particles within the CDM
halo virial radius and show the results as solid red curves. The error
bars we show in that figure represent the $1\sigma$ error in the
estimation of the average distribution while with dashed lines we plot
the unperturbed neutrino momentum distribution as given by equation
\ref{NU_pdf_vel}. In particular, the results have been extracted from
the simulations L60, L30, S60 and S30, for neutrinos with masses
$\Sigma_i m_{\nu_i}=0.60$ and 0.30 eV, respectively, and from L0 and
S30 for the CDM particles distribution.
 
We find that the neutrino velocity distribution within CDM haloes is,
for all CDM halo masses, closer to its unperturbed distribution for
lower neutrino masses. This is because the neutrino clustering becomes
smaller for lower neutrino masses. On the other hand, since the
neutrino clustering increases with the mass of its host CDM halo, the
deviation of the neutrino velocity distribution to its unperturbed
distribution increases for more massive CDM haloes. Furthermore, since
the clustering of neutrinos is larger for more massive neutrinos, the
velocity distribution deviates more from its unperturbed distribution
in the case of neutrinos with $\Sigma_i m_{\nu_i}=0.60$ eV (in
comparison with the 0.30 eV case). In all cases, we find that neutrino
velocities are, on average, larger than those from the CDM. The high
velocity tail is very well reproduced by the unperturbed velocity
distribution. This is not surprising since the dynamics of neutrinos
with large velocities is not strongly affected by gravity as can be
also seen from Fig.~\ref{Velocity_evolution}. These trends were also
obtained by \cite{Wong} when computing the neutrino momentum distribution
at the Earth neighborhood.

\section{Conclusions}
\label{Conclusions}

In this paper we have investigated the non-linear evolution of the
cosmic neutrino background.  We have used a modified version of {\sc
  GADGET}-3 that incorporates neutrinos as an independent particle
species. Since neutrino clustering is expected to be very small and
neutrino peculiar velocities are large, it is crucial to investigate
resolution, box-size effects and the initial redshift of the
simulations (as done by e.g. \cite{Brandbyge_haloes,Viel_2010,
  Bird_2011}).  We present quantitative results for the following
quantities: redshift evolution of the neutrino and CDM density fields
over the cosmological volume; redshift evolution of the neutrino and
CDM peculiar velocity fields over the cosmological volume; halo mass
function in neutrino cosmologies for a large range of halo masses;
neutrino properties inside virialized haloes (peculiar velocity and
density profiles); neutrino properties outside the virial radius for
isolated and not isolated haloes.

Our results can be summarized as follows.

\begin{itemize}
\item[-] In the cosmological volume the non-linear CDM and neutrino
  density fields evolve differently with cosmic time. Whereas the CDM
  density field evolve quickly, the neutrino density field evolves
  slowly and is mainly driven by the clustering of neutrinos within
  CDM halo potential wells.

\item[-] The neutrino momentum distribution within the cosmological
  volume deviates with respect to the unperturbed momentum
  distribution. These deviations increase with both $\Sigma_i
  m_{\nu_i}$ and $a=1/(1+z)$. At $z=0$, the fraction of neutrinos
  ($\Sigma_i m_{\nu_i}=0.60$ eV) with peculiar velocities smaller than
  100 km/s is a factor two smaller than the one predicted by the
  unperturbed momentum distribution.

\item[-] If we follow the neutrinos in momentum bins, neutrinos with
  $\Sigma_i m_{\nu_i}=0.60$ eV that have peculiar velocities of
  $\sim1/20$ and $\sim1/10$ of the mean cosmic peculiar velocity in
  the initial conditions start to behave like CDM at $z=3$ and $z=2$,
  respectively.

\item[-] We computed the halo mass function over four decades in mass
  and found a reasonably good agreement with the ST mass function by
  using $\Omega_{\rm M}=\Omega_{\rm{cdm}}+\Omega_{\rm b}$, without the
  $\Omega_{\nu}$ contribution. 

\item[-] We analysed the neutrino density profiles around CDM
  haloes. As expected, we found the presence of a large core for the
  most massive haloes above $10^{13.5}$ $h^{-1}$M$_\odot$: we provided
  a simple fitting formula, whose parameter values depend on neutrino
  mass, cold dark matter mass and redshift.

\item[-] For less massive haloes, the resolution of our N-body
  simulations do not allow us to probe the regime at which the core
  develops and thereby we present a simpler power-law fitting function
  to the density profile. For Milky Way size haloes, at $z=0$, the
  relic neutrino density at the solar radius would be enhanced, with
  respect to the neutrino background density, by more than a
  $\sim40\%$ for neutrinos with $\Sigma_i m_{\nu_i}=0.60$ and by more
  than $\sim10\%$ for neutrinos with $\Sigma_i m_{\nu_i}=0.30$.
  
\item[-] We also considered the peculiar velocity distribution of
  neutrino particles inside virialized haloes and compared this with
  the unperturbed Fermi-Dirac distribution. Important deviations take
  place in the low velocity tail: for neutrinos with $\Sigma_i
  m_{\nu_i}=0.60$ eV within CDM halos of masses $10^{14}$
  $h^{-1}$M$_\odot$, the fraction of those with velocities lower than
  $\sim200$ km/s, is more than a factor six smaller in the real
  distribution in comparison with the unperturbed distribution.
\end{itemize}

An accurate modelling and analysis of the impact of relic neutrinos on
cosmic structure in the non-linear regime is thus very important.  In
fact, strong and weak lensing observations of galaxy clusters are
already able to place constraints on the density profiles,
concentration and shape of these objects (see for example
\cite{Umetsu, Newman}). In this work we have shown that a cored
neutrino halo should be present around massive clusters and must impact
at some level on the overall cluster properties in a mass and redshift
dependent way.  Moreover, both spectroscopic and photometric surveys
of galaxies can probe the clustering of matter in a region that is
affected by the non-linearities of the neutrino component
(e.g. \cite{Xia}). Future large scale structure surveys like Euclid
\cite{Hamann,Audren,Euclid} are thus expected to place tight
constraints on neutrino properties by using clustering and weak
lensing observations of galaxies and galaxy clusters.

\section*{Acknowledgements}
Calculations for this paper were performed on SOM1 and SOM2 at IFIC and on the COSMOS Consortium supercomputer within the DiRAC Facility jointly funded by STFC, the Large Facilities Capital Fund of BIS and the University of Cambridge, as well as the Darwin Supercomputer of the University of Cambridge High Performance Computing Service (http://www.hpc.cam.ac.uk/),
provided by Dell Inc. using Strategic Research Infrastructure Funding
from the Higher Education Funding Council for England. We thank Volker
Springel for giving us permission to use {\sc GADGET}-3.  FVN is
supported by the ERC Starting Grant ``cosmoIGM''. SB is supported by
NSF grant AST-0907969 and the IAS. MV acknowledges support from
grants: INFN/PD51, ASI/AAE, PRIN MIUR, PRIN INAF 2009 ``Towards an
Italian Network for Computational Cosmology'' and from the ERC
Starting Grant ``cosmoIGM''. The authors wish to thank the referee for the report.

\appendix

\section{Impact of the CDM halo environment}
\label{Appendix_A}

In this paper we have studied the non-linear properties of relic
neutrinos within isolated CDM haloes. Here, we investigate how those
properties change when considering non-isolated CDM halos. In
Sec. \ref{neutrino_halo_sec} we defined a CDM halo as isolated if there were no more massive CDM halos located within a distance equal to 10 times the virial radius. If the former condition is not satisfied, then the CDM halo is non-isolated.

In Fig. \ref{enviroment} we show the average neutrino overdensity
profiles, at $z=0$, computed within isolated CDM haloes (red),
non-isolated CDM haloes (green) and within all CDM haloes (blue). The
masses of their host CDM haloes are equal to $10^{14}$
$h^{-1}$M$_\odot$ (upper left panel), $10^{13}$ $h^{-1}$M$_\odot$
(upper right panel), $10^{12}$ $h^{-1}$M$_\odot$ (bottom left panel)
and $10^{11}$ $h^{-1}$M$_\odot$ (bottom right panel) at $z=0$. Each
panel shows the overdensity profiles for two neutrino masses:
$\Sigma_i m_{\nu_i}=0.60$ eV (solid lines) and $\Sigma_i
m_{\nu_i}=0.30$ eV (dashed lines). In particular, the results shown have been extracted from the simulations L60 and L30, for CDM halo masses equal
or larger than $1\times10^{13}$ $h^{-1}$M$_\odot$, and from the
simulations S60 and S30 for CDM halo masses smaller than $1\times10^{13}$ $h^{-1}$M$_\odot$. Table \ref{env_table} shows the number of CDM haloes for each neutrino mass, each CDM halo mass and for each halo environment.

\begin{figure}
\begin{center}
\includegraphics[width=0.49\textwidth]{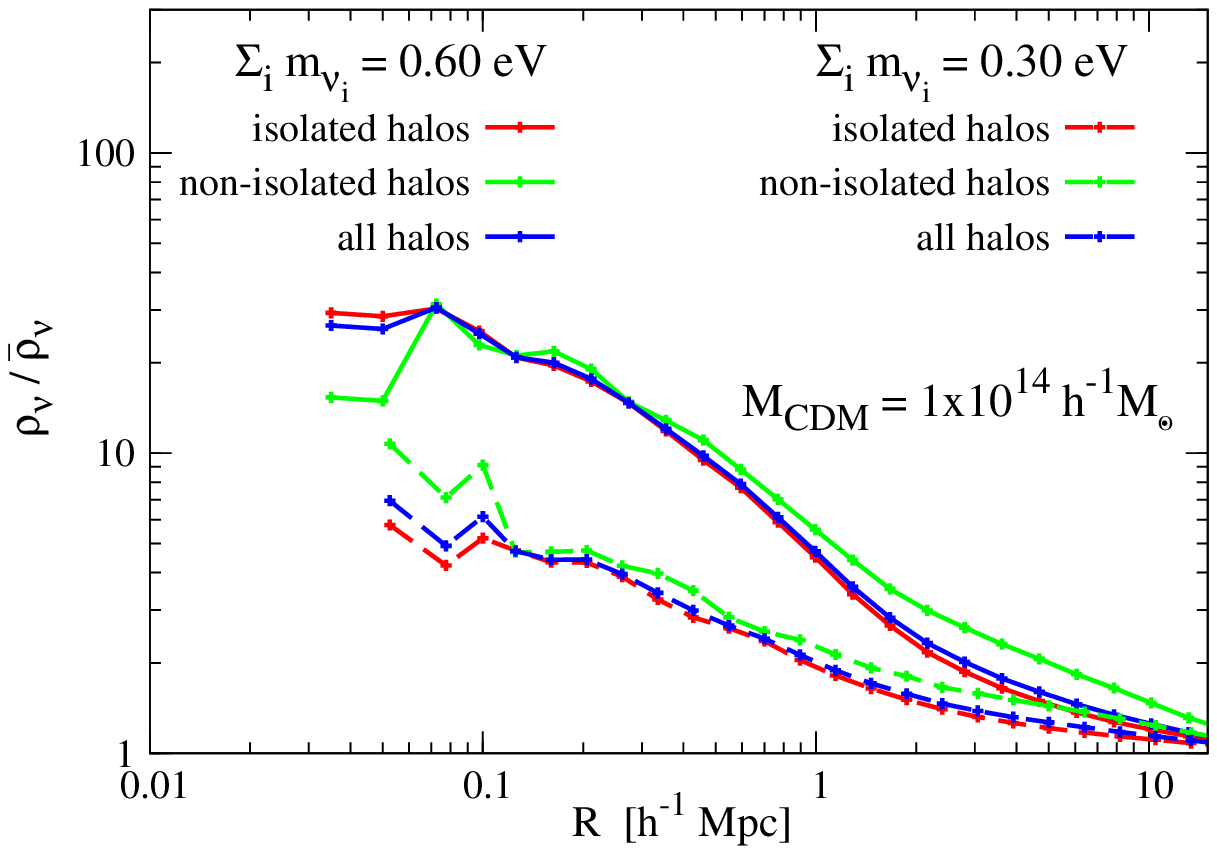}
\includegraphics[width=0.49\textwidth]{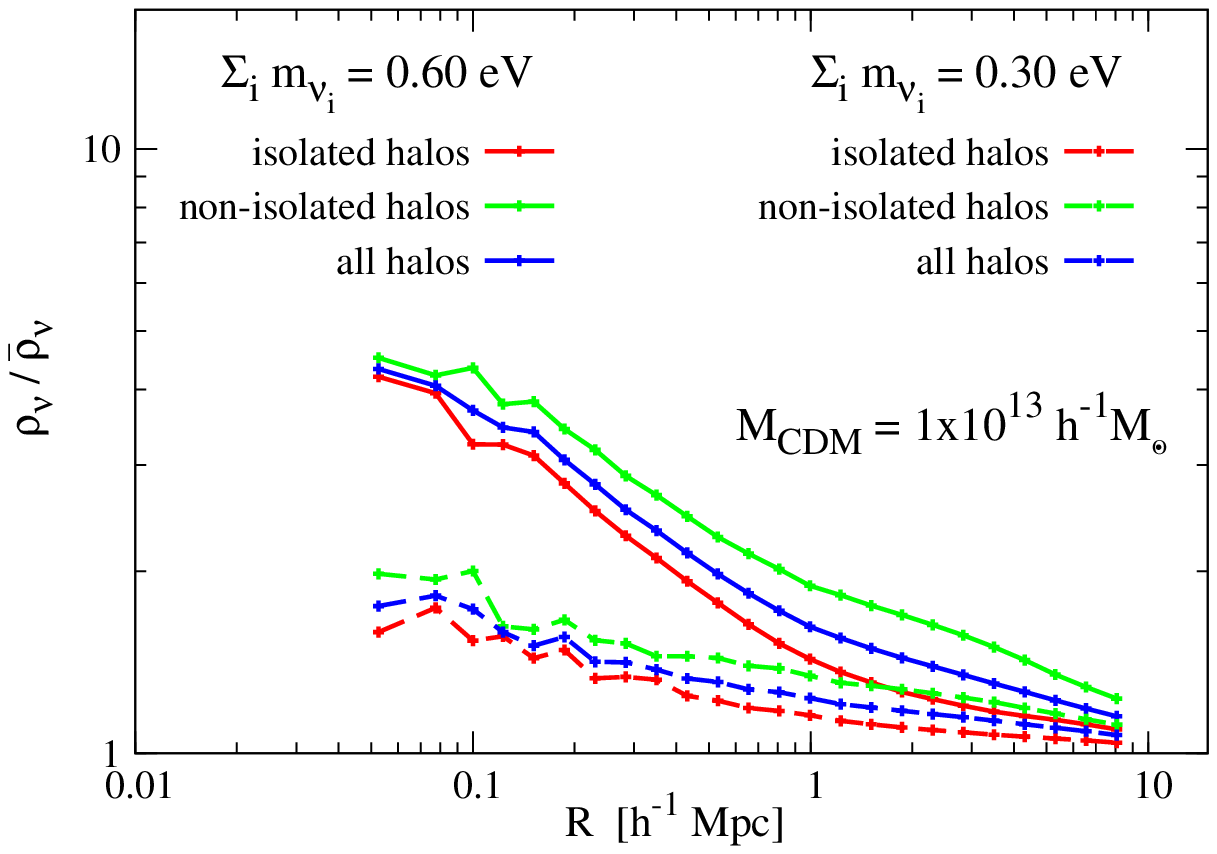}\\
\includegraphics[width=0.49\textwidth]{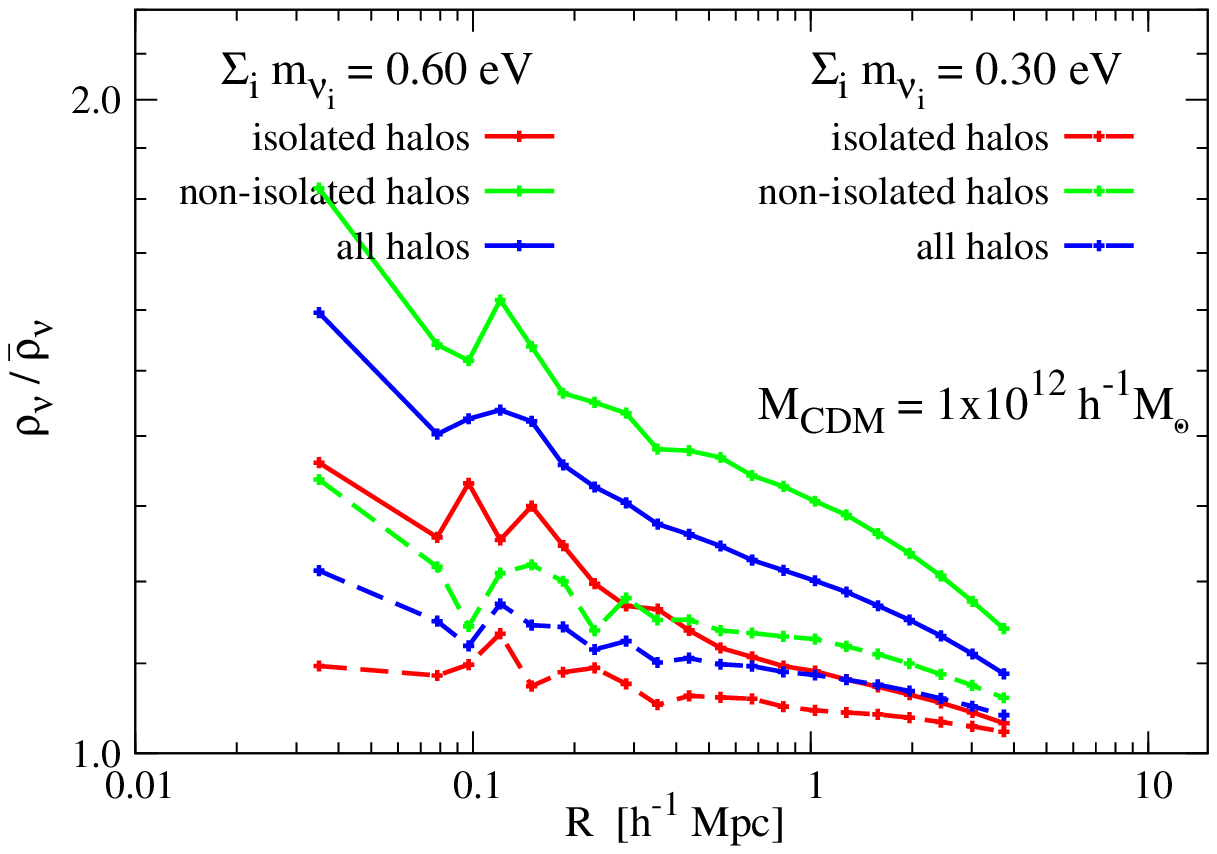}
\includegraphics[width=0.49\textwidth]{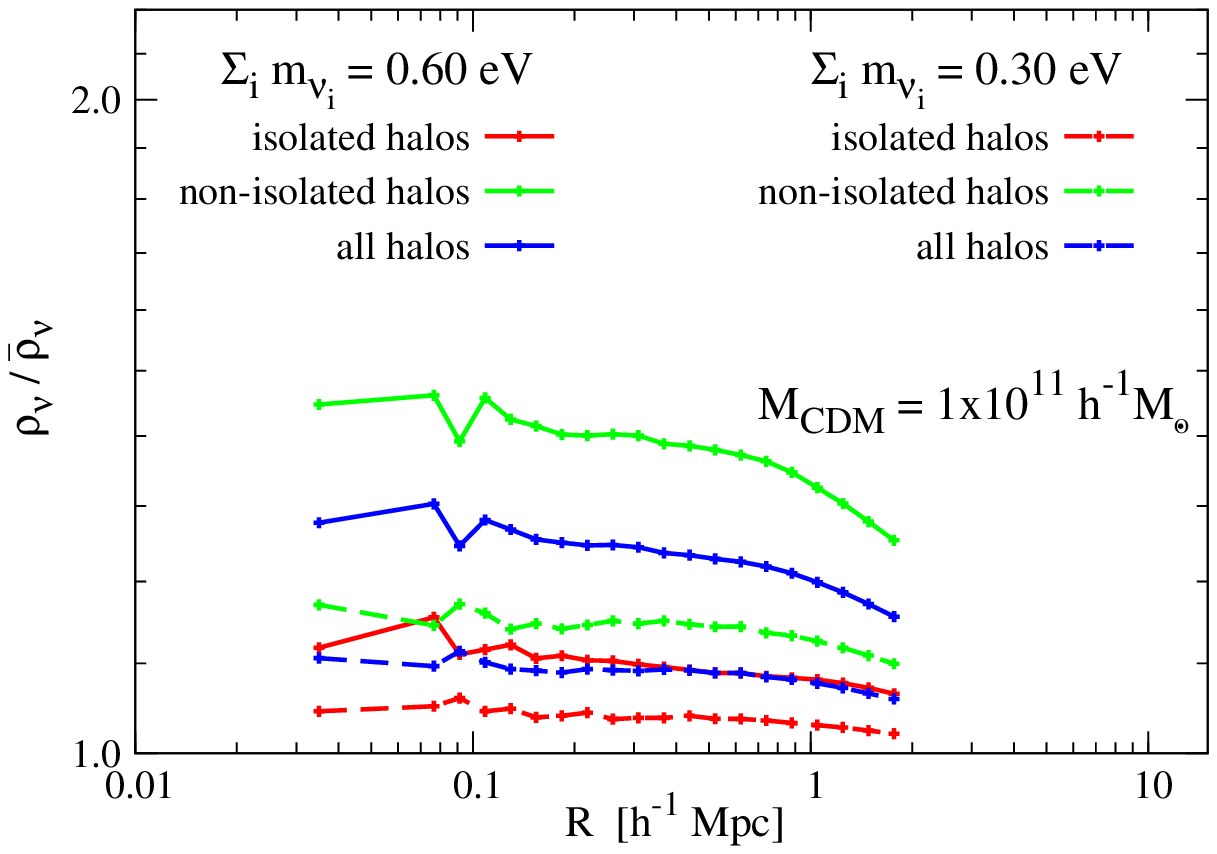}\\
\end{center}
\caption{Dependence of the average neutrino overdensity profiles with
  the environment of their host CDM haloes. The upper left panel shows
  the neutrino overdensity profiles within CDM haloes of masses equal
  to $10^{14}$ $h^{-1}$M$_\odot$, for neutrinos with $\Sigma_i
  m_{\nu_i}=0.60$ eV (solid lines) and $\Sigma_i m_{\nu_i}=0.30$ eV
  (dashed lines). The profiles are computed within isolated CDM haloes
  (red lines), non-isolated CDM haloes (green lines) and all haloes
  (blue lines). The other panels show the same for CDM halo masses
  equal to $10^{13}$ $h^{-1}$M$_\odot$ (upper right), $10^{12}$
  $h^{-1}$M$_\odot$ (bottom left) and $10^{11}$ $h^{-1}$M$_\odot$
  (bottom right).}
\label{enviroment}
\end{figure}

\begin{table}
\begin{center}
\begin{tabular}{|c||c|c|c|}
\hline
 & Isolated haloes & Non-Isolated haloes & All haloes \\
\hline
\hline 
$\Sigma_i m_{\nu_i}=0.60$ eV :  M$_{\rm{CDM}}=10^{14}$ M$_\odot$ & $205$ & $49$ & $254$ \\
\hline
$\Sigma_i m_{\nu_i}=0.30$ eV : M$_{\rm{CDM}}=10^{14}$ M$_\odot$ & $275$ & $86$ & $361$ \\
\hline
$\Sigma_i m_{\nu_i}=0.60$ eV : M$_{\rm{CDM}}=10^{13}$ M$_\odot$ & $2389$ & $1637$ & $4026$ \\
\hline
$\Sigma_i m_{\nu_i}=0.30$ eV : M$_{\rm{CDM}}=10^{13}$ M$_\odot$ & $2495$ & $1798$ & $4293$ \\
\hline
$\Sigma_i m_{\nu_i}=0.60$ eV : M$_{\rm{CDM}}=10^{12}$ M$_\odot$ & $171$ & $178$ & $349$ \\
\hline
$\Sigma_i m_{\nu_i}=0.30$ eV : M$_{\rm{CDM}}=10^{12}$ M$_\odot$ & $169$ & $160$ & $329$ \\
\hline
$\Sigma_i m_{\nu_i}=0.60$ eV : M$_{\rm{CDM}}=10^{11}$ M$_\odot$ & $1362$ & $1266$ & $2628$ \\
\hline
$\Sigma_i m_{\nu_i}=0.30$ eV : M$_{\rm{CDM}}=10^{11}$ M$_\odot$ & $1434$ & $1354$ & $2788$ \\
\hline
\end{tabular}
\end{center} 
\caption{Number of CDM haloes found in the simulations depending on
  their masses, on their environment and on the masses of the
  neutrinos. CDM haloes with masses equal or larger than $10^{13}$
  $h^{-1}$M$_\odot$ are extracted from the simulations L60 and L30,
  whereas the rest are extracted from the simulations S60 and S30.}
\label{env_table}
\end{table}

We find that the clustering of neutrinos is larger within non-isolated
CDM halos than within isolated CDM halos. This is not surprising since
we expect higher values in the density profile due to the
presence of a heavier halo in the vicinity of non-isolated CDM
haloes. The presence of the heavier halo may only locally modify the density proÞle, (e.g. when the halo is very far away), or it could inßuence the overall proÞle. The last situation corresponds to the
case in which, for example, the halo is a satellite of a much heavier
halo. In that case, the majority of the relic neutrinos will be
orbiting around the heavier halo, and therefore, the neutrino density
profile around the halo into study will be completely distorted since
it will be embedded into a larger and denser neutrino halo. As
expected, we find that the dispersion in the neutrino density profile
within non-isolated CDM haloes is larger than within isolated CDM
haloes.

It turns out that the fitting formulas \ref{VBV} and \ref{VBV_simple}
provide an excellent description of the neutrino density profiles,
independently of the environment of their host CDM haloes. The
environment of the CDM haloes affects, of course, the values of the
fitting formula parameters.

\begin{figure}
\begin{center}
\includegraphics[width=0.49\textwidth]{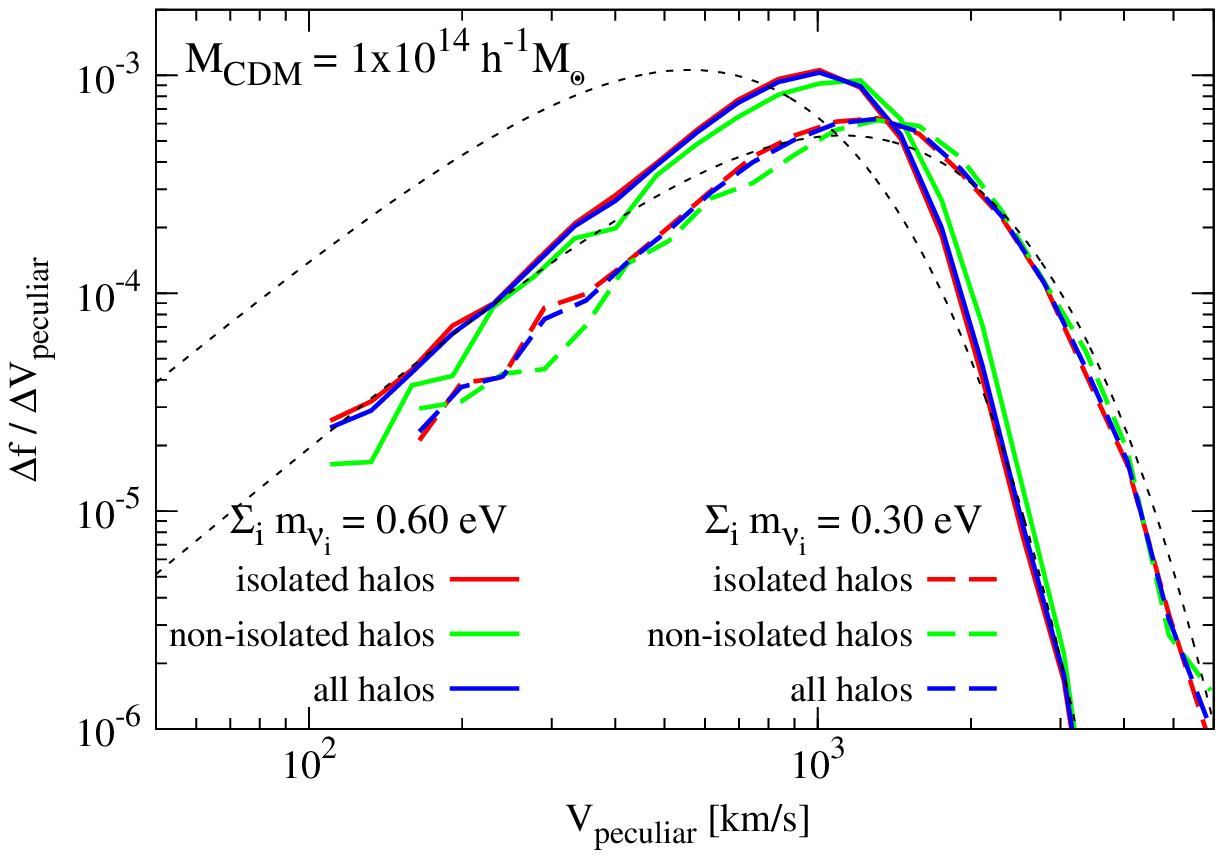}
\includegraphics[width=0.49\textwidth]{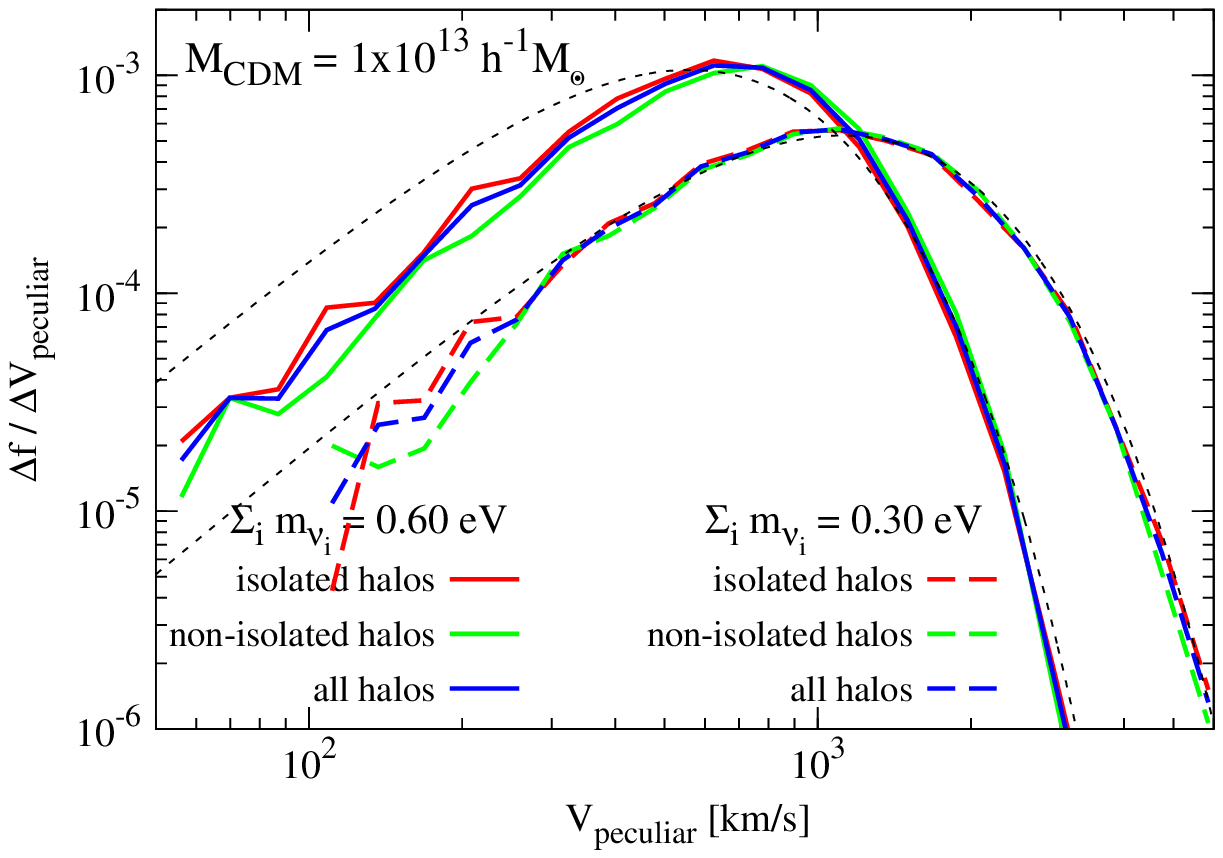}\\
\includegraphics[width=0.49\textwidth]{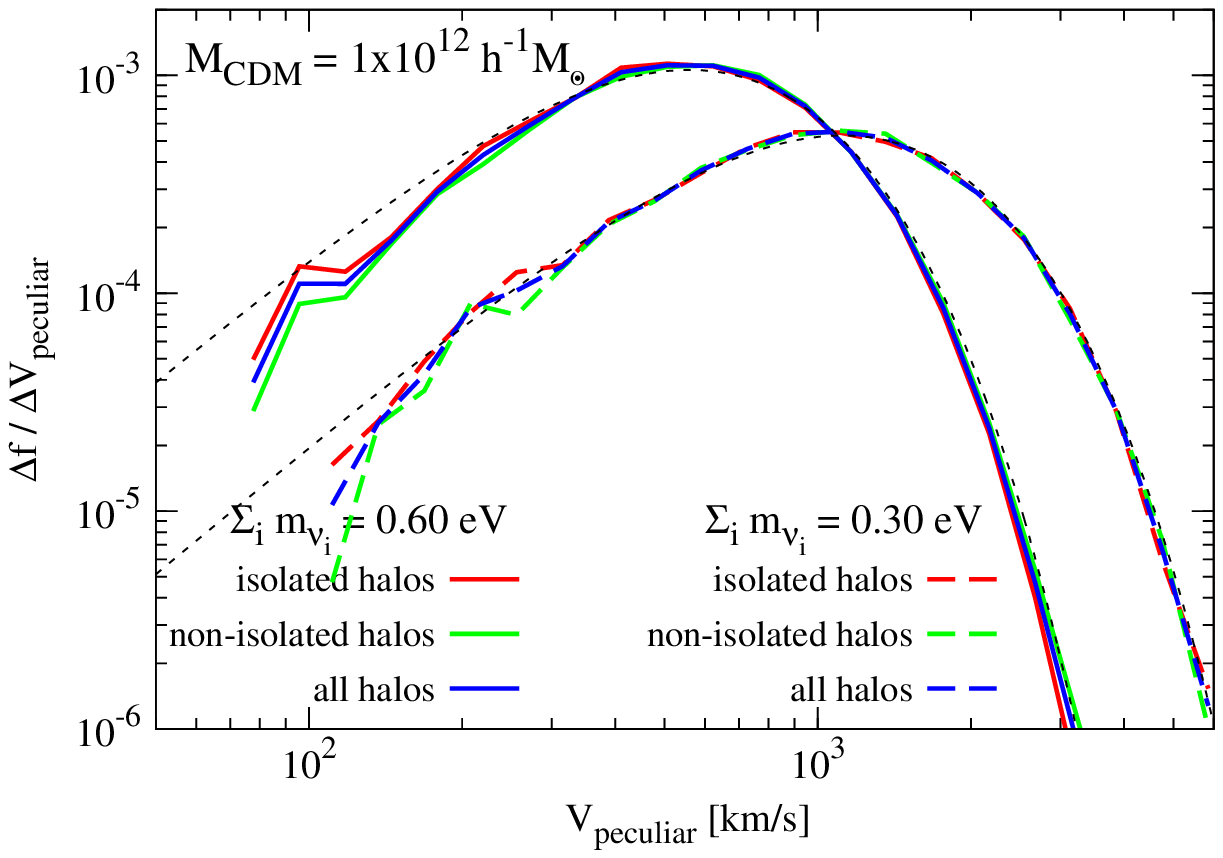}
\includegraphics[width=0.49\textwidth]{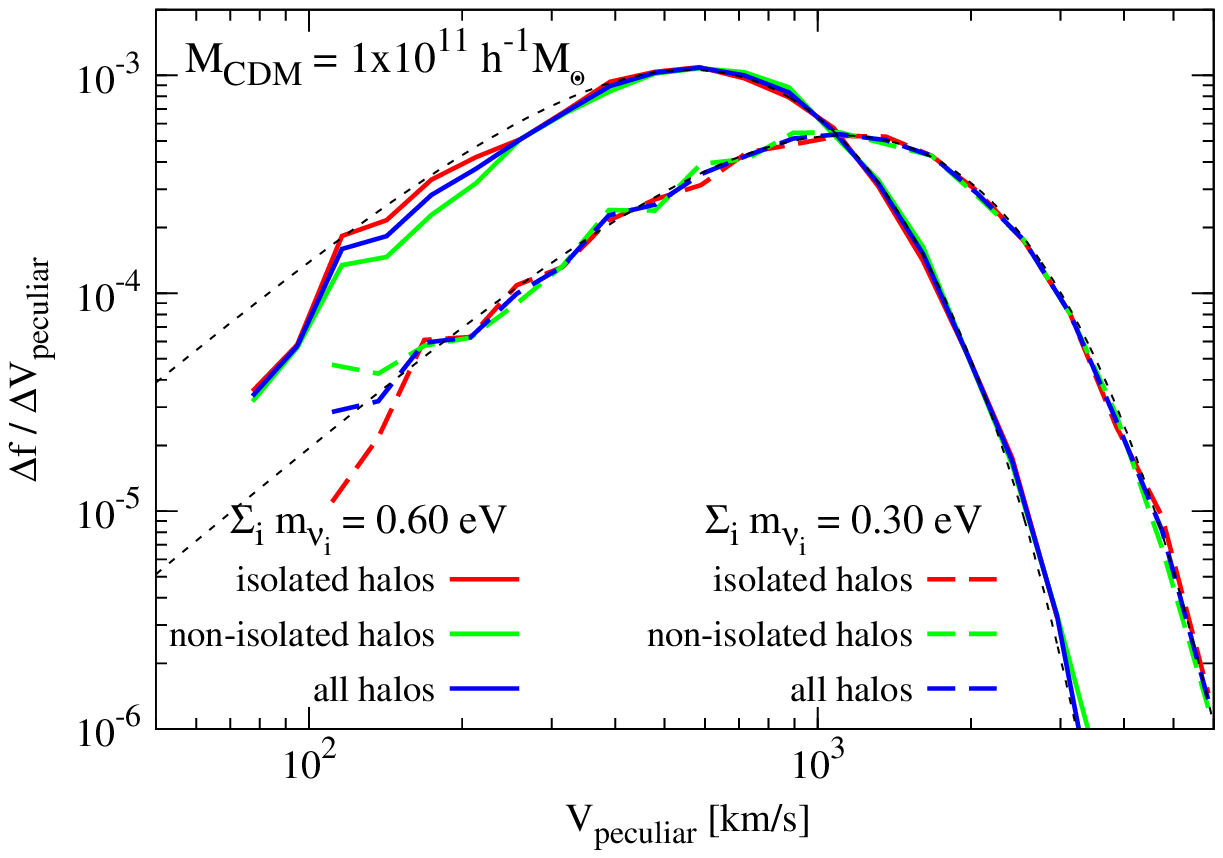}\\
\end{center}
\caption{Dependence of the distribution of the neutrino peculiar
  velocities with the environment of their host CDM halo. We show the
  proportion of neutrinos, within the virial radius of their host CDM
  haloes, with peculiar velocities between $V$ and $V+\triangle V$,
  per $\triangle V$, as a function of the neutrino peculiar velocity
  modulus for different CDM halo environments: isolated CDM haloes
  (red), non-isolated CDM haloes (green) and all CDM haloes
  (blue). Two neutrino masses, $\Sigma_i m_{\nu_i}=0.60$ eV (solid
  lines) and $\Sigma_i m_{\nu_i}=0.30$ eV (thick long-dashed lines), are studied
  within CDM haloes of masses equal to $10^{14}$ $h^{-1}$M$_\odot$
  (upper left panel), $10^{13}$ $h^{-1}$M$_\odot$ (upper right panel),
  $10^{12}$ $h^{-1}$M$_\odot$ (bottom left panel) and $10^{11}$
  $h^{-1}$M$_\odot$ (bottom right panel). The black thin short-dashed lines show
  the results of the unperturbed neutrino distribution as given by
  Eq. \ref{NU_pdf_vel}.}
\label{environment_PDF_vel_halo}
\end{figure}

Finally, in Fig. \ref{environment_PDF_vel_halo}, we investigate the impact of 
the CDM halo environment on the distribution of neutrino peculiar
velocities within the virial radius of CDM haloes. We examine a range of halo 
and neutrino masses in three different environments: isolated CDM haloes, 
non-isolated CDM haloes and all haloes.

Since we are focusing on the neutrino peculiar velocity distribution
within the CDM halo virial radius, it is expected that the CDM halo
environment does not play a critical role in the results. The presence
of a more massive CDM halo can only significantly modify the
distribution of neutrino velocities within the CDM halo virial radius
if it is close enough. If this is the case, then, as we find,
non-isolated CDM haloes will contain a smaller fraction of neutrinos
with low velocities. This is because the proportion of neutrinos with
large velocities, some of them belonging to a larger and deeper
neutrino halo centered in the more massive CDM halo, is enhanced for
non-isolated CDM halos due to the presence of a heavier CDM halo in
their neighborhood.

\section{Fitting formula parameters: dependence with $M_{\rm{CDM}}$ and $\Sigma_i m_{\nu_i}$}
\label{Appendix_B}

In table \ref{fitting} we show the preferred values for the parameters of equations \ref{VBV} and \ref{VBV_simple} over a wide range of CDM halo masses. They are presented for two neutrino masses: $\Sigma_i m_{\nu_i}=0.30$ eV and $\Sigma_i m_{\nu_i}=0.60$ eV. The parameter $\kappa$ in equation \ref{VBV_simple} is dimensionless. Thus for the above formulae the distance $r$ in \ref{VBV_simple} is assumed to be in $h^{-1}$ kpc.

\begin{table}
\begin{tabular}{cc|c|c|c|}
\cline{3-4}
& & \multicolumn{2}{c|}{$\Sigma_i m_{\nu_i}$} \\ \cline{3-4}
 & $z=0$ & 0.60 eV & 0.30 eV \\ \cline{1-4}
\multicolumn{1}{|c}{\multirow{3}{*}{${\rm M}\geqslant10^{13.5}$}} & \multicolumn{1}{|c|}{$\rho_c $} & $3.748\times10^{-8}{\rm M}^{0.64}$ & $6.056\times10^{-8}{\rm M}^{0.58}$\\ \cline{2-4}
\multicolumn{1}{|c}{} & \multicolumn{1}{|c|}{$r_c ~[h^{-1}{\rm kpc}]$} & $2.046\times10^{-4}{\rm M}^{0.43}$ & $4.029\times10^{-8}{\rm M}^{0.68}$\\ \cline{2-4}
\multicolumn{1}{|c}{} & \multicolumn{1}{|c|}{$\alpha$} & $-4.62+0.19~{\rm log}\left({\rm M}\right)$ & $-6.71+0.24~{\rm log}\left({\rm M}\right)$ \\ \cline{1-4}

\multicolumn{1}{|c}{\multirow{2}{*}{${\rm M}<10^{13.5}$}} & \multicolumn{1}{|c|}{$\kappa$} & $0.24+1.444\times10^{-20}{\rm M}^{1.7}$ & $0.19+3.242\times10^{-19}{\rm M}^{1.5}$   \\ \cline{2-4}
\multicolumn{1}{|c}{} & \multicolumn{1}{|c|}{$\alpha$} & $-3.64+0.15~{\rm log}\left({\rm M}\right)$ & $-2.06+0.209~{\rm log}\left({\rm M}\right)$ \\ \cline{1-4}
\end{tabular}
\caption{Formulae that reproduce the values of the parameters of the fitting formulas \ref{VBV} and \ref{VBV_simple} over a wide range of masses as shown in Fig. \ref{VBV_fig}. We have defined ${\rm M}={\rm M}_{\rm CDM}/(h^{-1}{\rm M}_\odot)$.}
\label{fitting}
\end{table}

\bibliographystyle{utcaps}
\bibliography{euclid}

\end{document}